\documentclass[letter,twocolumn]{autart}    

\usepackage[english]{babel}
\usepackage[normalem]{ulem}
\usepackage{graphics} 
\usepackage{epsfig} 
\usepackage{amsmath} 
\usepackage{amssymb}  
\usepackage{color}
\usepackage{graphicx}
\usepackage{algpseudocode}
\usepackage{algorithm}
\usepackage{natbib} 
\usepackage{url} 

\newtheorem{theorem}{Theorem}[section]
\newtheorem{definition}[theorem]{Definition}
\newtheorem{assumption}[theorem]{Assumption}
\newtheorem{lemma}[theorem]{Lemma}

\newtheorem{remark}[theorem]{Remark}
\newtheorem{example}[theorem]{Example}
\newtheorem{corollary}[theorem]{Corollary}
\newtheorem{proposition}[theorem]{Proposition}  
\newtheorem{problem}[theorem]{Problem}

\usepackage{enumitem}
\renewcommand{\theenumi}{\alph{enumi}}

\newcommand{\naturals}{\mathbb{N}}
\newcommand{\real}{\mathbb{R}}
\newcommand{\cplx}{\mathbb{C}}

\newcommand{\range}{\mathcal{R}}

\newcommand{\Dc}{\mathcal{D}}

\newcommand{\Fc}{\mathcal{F}}

\newcommand{\Ic}{\mathcal{I}}

\newcommand{\Kc}{\mathcal{K}}
\newcommand{\Lc}{\mathcal{L}}
\newcommand{\Mc}{\mathcal{M}}

\newcommand{\Pc}{\mathcal{P}}

\newcommand{\Vc}{\mathcal{V}}
\newcommand{\Wc}{\mathcal{W}}

\newcommand{\Pf}{\mathfrak{P}}

\newcommand{\EDMD}[3]{\operatorname{EDMD}(#1,#2,#3)}

\newcommand{\Kedmd}{K_{\operatorname{EDMD}}}

\newcommand{\until}[1]{\{1,\dots,#1\}}
\newcommand{\zuntil}[1]{\{0,\dots,#1\}}

\newcommand{\dx}{D(X)}
\newcommand{\dy}{D(Y)}

\newcommand{\tdx}{\tilde{D}(X)}
\newcommand{\tdy}{\tilde{D}(Y)}
\newcommand{\Span}{\operatorname{span}}

\newcommand{\re}{{\operatorname{re}}}
\newcommand{\im}{{\operatorname{im}}}

\newcommand{\rows}{{\operatorname{rows}}}
\newcommand{\cols}{{\operatorname{cols}}}
\newcommand{\rown}{{\operatorname{\sharp rows}}}
\newcommand{\coln}{{\operatorname{\sharp cols}}}
\newcommand{\basis}{{\operatorname{basis}}}

\newcommand{\ssd}{{\operatorname{SSD}}}
\newcommand{\tssd}{{\operatorname{T-SSD}}}

\newcommand{\cssd}{C_{\ssd}}
\newcommand{\dssd}{D_{\ssd}}
\newcommand{\dtssd}{D_{\tssd}}
\newcommand{\Kssd}{K_\ssd}
\newcommand{\Ktssd}{K_\tssd}
\newcommand{\ssddx}{\dssd(X)}
\newcommand{\ssddy}{\dssd(Y)}
\newcommand{\dtssdx}{\dtssd(X)}
\newcommand{\dtssdy}{\dtssd(Y)}
\newcommand{\ctssd}{C_{\tssd}}
\newcommand{\Xtest}{X_{\operatorname{test}}}
\newcommand{\Ytest}{Y_{\operatorname{test}}}

\newcommand{\symmintersection}{{\operatorname{Symmetric-Intersection}}}
\newcommand{\cmax}{C_{\max}}

\newcommand{\longthmtitle}[1]{\mbox{}{\textit{(#1):}}}
\newcommand{\setdef}[2]{\{#1 \; | \; #2\}}

\newcommand{\oprocendsymbol}{\hbox{$\square$}}
\newcommand{\oprocend}{\relax\ifmmode\else\unskip\hfill\fi\oprocendsymbol}

\allowdisplaybreaks

\graphicspath{{figures/}}

\begin{document}
\begin{frontmatter}

  \title{Generalizing Dynamic Mode Decomposition: Balancing Accuracy
    and Expressiveness in Koopman Approximations}

  \thanks{This work was supported by ONR Award N00014-18-1-2828 and
    NSF Award IIS-2007141.\newline A preliminary version of this paper
    appeared at the American Control Conference
    as~\citep{MH-JC:21-acc}.}
    
  \vspace{-10pt}
  
    \author[First]{Masih Haseli}%
    \author[First]{\quad Jorge Cort\'es}
  
    \address[First]{Department of Mechanical and Aerospace
      Engineering, University of California, San Diego,
      \{mhaseli,cortes\}@ucsd.edu}
          
\begin{abstract}
  This paper tackles the data-driven approximation of unknown
  dynamical systems using Koopman-operator methods.  Given a
  dictionary of functions, these methods approximate the projection of
  the action of the operator on the finite-dimensional subspace
  spanned by the dictionary.  We propose the Tunable Symmetric
  Subspace Decomposition algorithm to refine the dictionary, balancing
  its expressiveness and accuracy. Expressiveness corresponds to the
  ability of the dictionary to describe the evolution of as many
  observables as possible and accuracy corresponds to the ability to
  correctly predict their evolution.  Based on the observation that
  Koopman-invariant subspaces give rise to exact predictions, we
  reason that prediction accuracy is a function of the degree of
  invariance of the subspace generated by the dictionary and provide a
  data-driven measure to measure invariance proximity.  The proposed
  algorithm iteratively prunes the initial function space to
  identify a refined dictionary of functions that satisfies the
  desired level of accuracy while retaining as much of the original
  expressiveness as possible. We provide a full characterization of the
  algorithm properties and show that it generalizes both Extended
  Dynamic Mode Decomposition and Symmetric Subspace
  Decomposition. Simulations on multiple systems show the effectiveness
  of the proposed methods in producing Koopman approximations of
  tunable accuracy that capture relevant information about the
  dynamical system.
\end{abstract}

\begin{keyword}
  Nonlinear system identification, Koopman operator, Dynamic Mode
  Decomposition, accurate prediction
\end{keyword}
  
\end{frontmatter}

\section{Introduction}\label{sec:introduction}
Progress in data acquisition and labeling, along with widespread
access to high-performance computing capabilities for storing,
processing, and data analysis, has resulted in a surge of activity in
learning and modeling of dynamical phenomena across multiple
domains. In this context, the importance of identification techniques
that yield tractable representations of nonlinear dynamics rooted at a
solid theoretical framework cannot be overemphasized. One such
technique is Koopman operator theory which, instead of prescribing the
evolution of system trajectories as state-space models do, describes
the evolution of functions defined over the state space
(a.k.a. observables). The infinite-dimensional nature of the operator
has prevented its widespread use due to the lack of computational
methods to represent it. Extended dynamic mode decomposition (EDMD)
addresses this by employing data from the dynamics to approximate the
projection of the action of the Koopman operator on a
finite-dimensional subspace spanned by a predefined dictionary of
functions.

Despite EDMD's success, it is still not well understood how to choose
dictionaries that both capture relevant information about the dynamics
and are able to accurately predict its evolution.  Prediction accuracy
is related to the degree of invariance, with respect to the operator,
of the subspace generated by the dictionary and, in fact, can be
improved by selectively pruning its functions.  Such process, however,
impacts expressiveness, understood as the ability of the dictionary to
describe the evolution of as many observables as possible. This paper
is motivated by the need to address the accuracy-expressiveness
trade-off in dictionary~selection.

\emph{Literature Review:} Given a dynamical system, its associated
Koopman operator~\citep{BOK:31,BOK-JVN:32} is a linear operator
characterizing the effect of the dynamics on functions in a (generally
infinite-dimensional) linear function space. The values of its
eigenfunctions also evolve linearly in time on the trajectories of the
system.  These properties enable the use of the spectral properties of
the operator to process data from dynamical systems using efficient
linear algebraic methods fully compatible with the arithmetic
operations of digital computers~\citep{IM:05,MB-RM-IM:12}. This has
led to many applications in fluid
dynamics~\citep{CWR-IM-SB-PS-DSH:09}, model reduction~\citep{IM:05},
and control, including controller
synthesis~\citep{HC-UV-YC:20,CF-YC-ADM-JWB:20,DG-DAP:21}, model
predictive control~\citep{MK-IM-automatica:18,SHS-AN-JSK:20}, control
of PDEs~\citep{SP-SK:17}, and robotic
applications~\citep{GM-MC-XT-TM:19,VZ-EB:22}.

The infinite-dimensional nature of the Koopman operator is an
impediment to its direct implementation on digital computers.  A
popular way to find finite-dimensional representations of the operator
is through Dynamic Mode Decomposition (DMD) and its variants,
initially developed to extract dynamical information from data about
fluid flows~\citep{PJS:10,JHT-CWR-DML-SLB-JNK:13}.
\citet{MOW-IGK-CWR:15} developed the Extended Dynamic Mode
Decomposition (EDMD) algorithm, a variant of DMD capable of
approximating the projection of the action of the Koopman operator
from data on a finite-dimensional space spanned by a chosen dictionary
of functions. \citet{MK-IM:18} formally established the convergence of
the EDMD approximation to the projection of the action of the operator
on the span of the dictionary. Recently, \citet{HL-DMT:20}
  analyzed the accuracy of long-term prediction by DMD and its
  variants. \citet{GM-MLC-XT-TDM:21} used a Taylor expansion method to
  enrich the dictionary for EDMD to achieve lower errors in long-term
  predictions. \citet{CZ-EZ:21} used finite element methods to learn
  Koopman approximations with accuracy bounds quantifying their
  quality. This work also provided a variant of EDMD to learn the
  Koopman generator combined with finite element
  methods. \citet{FN-SP-FP-MS-KW:21} provided several probabilistic
  bounds for approximation of the Koopman operator based on the
  availability of only finitely many data points both for
  deterministic and stochastic systems.  It is worth mentioning
  that (E)DMD captures valuable information about stochastic dynamical
  systems~\citep{MOW-IGK-CWR:15,SK-FN-SP-JHN-CC-CS:20}; however, it is
  sensitive to the existence of measurement noise in the available
  data. \citet{STMD-MSH-MOW-CWR:16} and \citet{MH-JC:19-acc} provide
  methods to deal with measurement noise in data for DMD and EDMD,
  respectively.

  In general, given an arbitrary dictionary, there is no guarantee
  that EDMD provides an accurate approximation for all the observables
  in the span of the dictionary.  This has resulted in the search for
  dictionaries that span invariant
  subspaces~\citep{SLB-BWB-JLP-JNK:16} under the Koopman operator, on
  which the EDMD approximation is exact. The work \citep{CAJ-EY:18}
  introduces a class of logistic functions to approximate
  Koopman-invariant subspaces for systems whose dynamics are known.
  On the other hand, given unknown systems and using data sampled from
  their trajectories, Koopman-invariant subspaces are approximated
  using neural networks
  in~\citep{NT-YK-TY:17,BL-JNK-SLB:18,EY-SK-NH:19,SEO-CWR:19} and
  sparsity-promoting methods in~\citep{SP-NAM-KD:21}. The works
  by~\citep{EK-JNK-SLB:21,MK-IM:20} directly approximate Koopman
  eigenfunctions, which naturally span Koopman-invariant subspaces.
  These data-driven methods do not provide theoretical guarantees for
  the resulting approximations. Given the importance of such
  guarantees, our previous work~\citep{MH-JC:22-tac,MH-JC:21-tcns} has
  provided necessary and almost surely sufficient conditions for the
  identification of the maximal Koopman-invariant subspace and all
  Koopman eigenfunctions in an arbitrary finite-dimensional function
  space. We have also provided approximations to identify subspaces
  that are close to being invariant for cases when the maximal
  Koopman-invariant subspace does not capture enough information about
  the dynamics.

  It is important to note that the existence of finite-dimensional
  Koopman-invariant subspaces containing the states of the system is
  not guaranteed~\citep{SLB-BWB-JLP-JNK:16}. However, invariant
  subspaces still contain Koopman eigenfunctions, and can capture
  relevant information about the vector field, physical constraints,
  conservation laws, stability, and even the construction of Lyapunov
  functions, see e.g.,~\citep{AM-IM:16}.  Nonetheless, in some
  applications, one can tolerate a certain level of inaccuracy in
  order to capture a more diverse function space that is not
  necessarily Koopman invariant but captures important variables such
  as the states of the system.

\begin{figure}[tbh]
  \centering%
  \includegraphics[width=.75\linewidth]{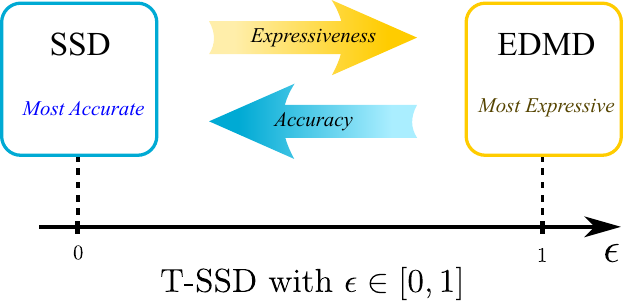}
  \caption{T-SSD generalizes SSD and EDMD.  Given an
      \emph{arbitrary} finite-dimensional function space, by
    changing the design parameter~$\epsilon \in [0,1]$ in T-SSD, one
    can strike a balance between the invariance proximity of the
    identified subspace (i.e., the accuracy of the resulting model
    based on the available data) and its
    expressiveness.}\label{fig:T-SSD-generalization}
\end{figure}

\emph{Statement of Contributions:} We consider the problem of
  data-driven identification of finite-dimensional spaces that are
  close, with tunable accuracy, to being invariant under the action of
  the Koopman operator. Our main result, illustrated in
  Figure~\ref{fig:T-SSD-generalization}, consists of the synthesis of
  a computational procedure, termed Tunable Symmetric Subspace
  Decomposition (T-SSD), that given an \emph{arbitrary}
  finite-dimensional function space, balances the trade-off between
  the expressiveness\footnote{We note that notions of
      expresiveness different that the one adopted here are also
      possible. For instance, and although out of the scope of the
      paper, expresiveness could also be understood as the ability of
      the dictionary to describe specific finitely many predefined
      observables of interest, such as state variables, in a
      \emph{given} set of coordinates.} of its subspaces and the
  accuracy of the Koopman approximations on them.

The roadmap of supporting contributions leading to the design and full
characterization of T-SSD is as follows.  Our first contribution
builds on the observation that the proximity of a function space to
being invariant is a measure of its (and consequently its
  members') prediction accuracy under finite-dimensional Koopman
approximations, as an exact invariant subspace leads to exact
predictions of the evolution of observables.  We introduce the novel
notion of $\epsilon$-apart spaces to measure invariance proximity
using data snapshots sampled from the trajectories of the
unknown dynamics.  Using this notion, and given an arbitrary
finite-dimensional function space spanned by a dictionary of
functions, we formulate our objective as that of finding a parametric
family of subspaces whose value of the parameter determines the
desired level of invariance proximity.  This parametric family can be
viewed as balancing invariance proximity (i.e., prediction accuracy)
and the dimension of the subspace (i.e., expressiveness).  

Given a desired accuracy parameter, our second contribution is the
design of T-SSD as an algorithmic procedure that finds a function
space satisfying the desired accuracy by iteratively removing the
functions in the span of the original dictionary that violate the
desired accuracy.  We show that T-SSD terminates in a finite number of
iterations and characterize its computational complexity.  Moreover,
we show that its identified subspaces contain the maximal
Koopman-invariant subspace and all Koopman eigenfunctions in the span
of the original dictionary.  We also show that the accuracy parameter
bounds the relative root mean square prediction error for all
(uncountably many) functions in the identified subspace.  This
advantage of the T-SSD algorithm in deriving accuracy bounds on the
prediction of individual functions independently of linear changes of
coordinates stems from focusing on the subspaces instead of their
basis.  Our next contribution establishes that both Extended Dynamic
Mode Decomposition and Symmetric Subspace Decomposition algorithms are
particular cases of T-SSD, cf. Figure~\ref{fig:T-SSD-generalization}.

Our final contribution is a computationally efficient version of T-SSD
with drastically lower computational complexity when the number of
data snapshots is significantly larger than the dimension of the
original dictionary.  We illustrate in simulation the effectiveness of
T-SSD in identifying informative Koopman approximations of tunable
accuracy.

\section{Preliminaries}\label{sec:preliminaries}

Here, we introduce the notation used in the paper and provide a
  brief overview of Koopman operator theory, extended dynamic mode
  decomposition (EDMD), and symmetric subspace decomposition (SSD).

\subsection{Notation}
We let $\real$, $\cplx$, $\naturals_0$, and $\naturals$
represent the sets of real, complex, nonnegative integer, and
natural numbers respectively.  Given a matrix $A\in \cplx^{m \times
	n}$, we denote its transpose, conjugate transpose, pseudo-inverse,
range space, and Frobenius norm by $A^T$, $A^H$, $A^\dagger$,
$\range(A)$, and $\|A\|_F$ respectively. In addition, $\cols(A)$,
$\rows(A)$, $\coln(A)$, and $\rown(A)$ represent its set of columns,
set of rows, number of columns, and number of rows. Also if $A$ is a
nonsingular square matrix, $A^{-1}$ denotes its inverse. Given $A
\in \cplx^{m \times n}$ and $B \in \cplx ^{m \times p}$, we use
$[A,B]$ to represent the matrix formed by their side by side
concatenation. We denote by $\mathbf{0}_{m \times n}$ and $I_n$, the
$m \times n$ zero matrix and the identity matrix of size $n$
respectively (we omit the indices when the context is clear). Given
a vector $v \in \cplx^n$, $\|v\|_2 := \sqrt{v^H v}$ denotes its
2-norm. 

We denote by $\Span\{v_1,\ldots, v_k \}$ the vector space over $\cplx$
spanned by $v_1,\ldots, v_k \in \cplx^n$. Given functions
$f_1,\ldots, f_k $, $\Span\{f_1,\ldots, f_k \}$ is the function
space formed by all functions in the form of $c_1f_1+\cdots+c_kf_k$
with $\{c_i\}_{i=1}^k \subset \cplx$. For a vector space
$\Vc \subseteq \real^m$, $\Pc_{\Vc}$ denotes the orthogonal projection
operator on $\Vc$. For convenience, we denote the orthogonal
projection operator on the range space of a matrix $A$ by~$\Pc_A$,
which takes the form $\Pc_A \,w = A A^\dagger w$, for $w \in \real^m$.
For vectors $v, w \in \real^m$, $v \perp w$ indicates that $v$ and $w$
are orthogonal. Given vector spaces $\Vc_1, \Vc_2 \subseteq \real^m$,
$\Vc_1 \perp \Vc_2$ denotes that the vector spaces are orthogonal,
i.e., all vectors in $\Vc_1$ are orthogonal to all vectors in
$\Vc_2$. We define the sum of vector spaces $\Vc_1,\Vc_2$ by
$\Vc_1+\Vc_2 := \{v_1+v_2| v_1 \in \Vc_1 \land v_2 \in \Vc_2\}$.
Given sets $S_1, S_2$, we denote their union and intersection by
$S_1 \cup S_2$ and $S_1 \cap S_2$ respectively.  Given functions
$f: S_2 \to S_3$, $g: S_1 \to S_2$, $f \circ g: S_1 \to S_3$ denotes
their composition.

\subsection{Koopman Operator}
Our exposition follows~\citet{MB-RM-IM:12}: given the discrete-time
dynamics
\begin{align}\label{eq:dymamical-sys}
  x^+ = T(x).
\end{align}
defined on the state space $\Mc \subseteq \real^n$ and a linear space
of functions $\Fc$ defined over the field $\cplx$ and closed under
composition with $T$, the Koopman operator $\Kc: \Fc \to \Fc$
associated with~\eqref{eq:dymamical-sys} is defined as
$ \Kc f = f \circ T$.  The functions in $\Fc$ are called
\emph{observables}. As a consequence of the linearity of $\Fc$, the
Koopman operator is \emph{spatially linear}, i.e.,
\begin{align}\label{eq:Koopman-spatial-linear}
  \Kc (c_1 f_1+ c_2 f_2) = c_1 \Kc(f_1) + c_2 \Kc(f_2),
\end{align}
for any $f_1,f_2 \in \Fc$ and $c_1,c_2 \in \cplx$. The linearity of
the Koopman operator paves the way for defining its
eigendecomposition. Formally, a function $\phi \in \Fc$ is an
\emph{eigenfunction} of the Koopman operator with \emph{eigenvalue}
$\lambda \in \cplx$ if
\begin{align}\label{eq:Koopman-eigendecomposition}
  \Kc\phi = \lambda \phi.
\end{align}

The eigenfunctions of the Koopman operator evolve linearly in time on
the trajectories of~\eqref{eq:dymamical-sys},
\begin{align}\label{eq:eig-linear-evolution}
  \phi(x^+) =\phi \circ T(x) = \Kc\phi (x) = \lambda \phi(x).
\end{align}

The linear temporal evolution of eigenfunctions combined
with~\eqref{eq:Koopman-spatial-linear} make the Koopman operator a
powerful tool for \emph{linear prediction} of the functions' values on
the trajectories of the (generally nonlinear) dynamical
system~\eqref{eq:dymamical-sys}. Formally, given a function $f
=\sum_{i=1}^{N_k} c_i \phi_i$, where $\{c_i\}_{i=1}^{N_k} \subset
\cplx$ and $\{\phi_i\}_{i=1}^{N_k}$ are eigenfunctions of $\Kc$ with
eigenvalues $\{\lambda_i\}_{i=1}^{N_k}$, one can predict the values of
$f$ on the trajectory $\{x(j)\}_{j \in \naturals_0}$ starting from initial condition $x(0) \in \Mc$ as
\begin{align}\label{eq:function-evolution-Koopman}
  f(x(j)) = \sum_{i=1}^{N_k} c_i \lambda_i^j \, \phi_i (x(0)), \quad
  \forall j \in \naturals_0.
\end{align}
Finally, a subspace $\Lc \subseteq \Fc$ is \emph{Koopman-invariant} if
for every $f \in \Lc$, we have $\Kc f \in \Lc$. Trivially, any
subspace spanned by Koopman eigenfunctions is Koopman invariant.

\subsection{Extended Dynamic Mode Decomposition}

In general, the Koopman operator is infinite dimensional. Moreover, in
many practical data-driven applications, the dynamical system is
unknown and only data from some trajectories is available. These
issues motivate the use of data-driven methods to approximate the
effect of the Koopman operator on finite-dimensional subspaces, as we
discuss next.  Here, we recall the Extended Dynamic Mode Decomposition
(EDMD) method following~\citet{MOW-IGK-CWR:15}.  EDMD uses data from
the trajectories of the system to approximate the action of the
Koopman operator on a predefined function space. The first
ingredient of EDMD is the data matrices $X,Y \in \real^{N \times n}$
containing $N$ data snapshots, where corresponding rows of $X,Y$
characterize two consecutive points on a trajectory of the
system. Formally,
\begin{align}\label{eq:data-snapshots}
  y_i = T(x_i), \, \forall i \in \until{N},
\end{align}
where $x_i^T$ and $y_i^T$ are the $i$th rows of $X$ and $Y$
respectively.  The second ingredient of EDMD is a dictionary $D: \Mc
\to \real^{1 \times N_d}$ of $N_d$ functions, denoted as
\begin{align}\label{eq:dictionary-def}
  D(x) = [d_1(x), \ldots, d_{N_d}(x)],
\end{align}
where $d_i: \Mc \to \real$ for all $i \in \until{N_d}$. The dictionary
spans a finite-dimensional space of functions over $\cplx$, and its
elements can be complex-valued functions in general. However, since
the system is defined over the state space $\Mc \subset \real^n$,
complex Koopman eigenfunctions form complex conjugate pairs which can
be fully represented by a pair of real-valued functions. For this
reason, and without loss of generality, we use real-valued functions
as dictionary elements.

EDMD formulates a least-squares optimization to find the best fit for
the linear evolution of the dictionary functions on the data. If we
denote the effect of the dictionary on an arbitrary data matrix
$Z \in \real^{N \times n}$ by $D(Z) = [D(z_1)^T, \ldots, D(z_N)^T]^T$,
where $z_i^T$ corresponds to the $i$th row of $Z$, then one can write
the EDMD's optimization problem in compact form as
\begin{align}\label{eq:EDMD-optimization}
  \underset{K}{\text{minimize}} \| D(Y) - D(X) K \|_F.
\end{align}
whose closed-form solution is
\begin{align}\label{eq:EDMD-closed-form}
  \Kedmd = \EDMD{D}{X}{Y} := \dx^\dagger \dy.
\end{align}
$\Kedmd$ approximates the projection of the action of the Koopman
operator on $\Span(D)$ as follows. Under the identification of
$\cplx^{N_d}$ with $\Span (D)$ defined by $v \leftrightarrow D(\cdot) v$, 
this approximation takes the form
\begin{align}\label{eq:linear-map}
  v \mapsto \Kedmd v .
\end{align}
As a result, for the function
  $f (\cdot) = D(\cdot) v_f \in \Span(D)$, one can define the EDMD's
  approximation of $\Kc f$ as
\begin{align}\label{eq:function-predictor}
\Pf_{\Kc f} = D(\cdot) \Kedmd v_f.
\end{align}
It is important to note that $\Pf_{\Kc f}$ can be viewed as the
$L_2$-orthogonal projection of $\Kc f$ on $\Span(D)$ given an
empirical measure defined based on the rows of
$X$~\citep{SK-PK-CS:16,MK-IM:18}.  Moreover, the eigendecomposition of
$\Kedmd$ provides insight into the eigendecomposition of the Koopman
operator. Formally, given an eigenpair $(\lambda,v)$ of $\Kedmd$, one
approximates an eigenfunction of the Koopman operator with eigenvalue
$\lambda$ as
\begin{align}\label{eq:EDMD-eigenfunction}
  \phi(\cdot) := D(\cdot) v.
\end{align}
Note that the EDMD predictor in~\eqref{eq:function-predictor} applied
to the eigenfunction $\phi$ in~\eqref{eq:EDMD-eigenfunction} leads to
$\Pf_{\Kc \phi} = \lambda \phi$, which is in agreement with the linear
evolution~\eqref{eq:eig-linear-evolution} of Koopman eigenfunctions.
Note that EDMD provides $N_d$ Koopman (generalized) eigenfunction
approximations even if the Koopman operator does not have $N_d$
eigenfunctions in the span of the dictionary.  If the dictionary $D$
spans a Koopman-invariant subspace, then EDMD completely captures the
behavior of the Koopman operator (and all its eigenfunctions) on
$\Span(D)$.  As a result, $\| \dy - \dx\Kedmd\|_F=0$,
and~\eqref{eq:function-predictor} provides exact prediction,
independently of the data used for EDMD's training.
  
We refer the reader to~\citep{MK-IM:18} for \emph{asymptotic}
convergence results as $N_d \rightarrow \infty$, concerning $\Kedmd$
capturing the spectrum of the Koopman operator as well as the
finite-horizon prediction of EDMD for observables in $\Span(D)$.  It
is essential to keep in mind that the aforementioned asymptotic
convergence results do not imply that a larger dictionary is
necessarily closer to being invariant under the Koopman operator. In general, asymptotic behavior requires a
\emph{sufficiently large} dictionary; however, unless other properties
(e.g., monotonicity) exist, the dictionary's quality (in terms of prediction accuracy) might deteriorate
by adding more functions. We illustrate this point in the following
example.
  
\begin{example}\longthmtitle{A Larger Dictionary is Not Necessarily
    Better}\label{ex:more-not-better} 
  {\rm Consider the linear system $x^+ = 0.5 x$ with state space
    $\Mc = \real$. Consider the dictionaries, $D_1(x) = [x]$ and
    $D_2(x) = [ x, x^3-x^2]$. Clearly,
    $\Span(D_1) \subsetneq \Span(D_2)$. However, $\Span(D_1)$ is
    invariant under the Koopman operator and EDMD's prediction (which
    coincides with the dynamics) is exact. On the other hand, despite
    being a larger dictionary (and containing $D_1$), EDMD with
    dictionary $D_2$ does not lead to accurate predictions on
    $\Span(D_2)$. The reason for this issue is that by adding the
    function $d(x) = x^3 -x^2$ to dictionary $D_1$, we break the
    invariance of its span, leading to a larger space $\Span(D_2)$
    which is not invariant.  \oprocend }
\end{example}
  
A final note regarding EDMD is that it is not designed to work with
data corrupted with measurement noise. Hence, data pre-processing
might be needed to take full advantage of the methods proposed in the
paper, which rely on~EDMD.

\subsection{Symmetric Subspace Decomposition}\label{sec:SSD}
In general, since the true system dynamics is unknown, choosing a
dictionary that spans a Koopman-invariant subspace is
challenging. This justifies the importance of developing data-driven
methods that aid in this task.  Here, we recall the Symmetric Subspace
Decomposition (SSD) algorithm following~\citet{MH-JC:22-tac}. Starting
from a dictionary $D$ and data snapshots $X,Y$, the SSD algorithm,
cf. Algorithm~\ref{algo:ssd}, finds a basis for the maximal
Koopman-invariant subspace in $\Span(D)$. To achieve this goal, SSD
relies on the following.

\begin{assumption}\longthmtitle{Full Rank Dictionary
    Matrices}\label{a:full-rank}
  The matrices $\dx$ and $\dy$ have full column rank.  \oprocend
\end{assumption}

Assumption~\ref{a:full-rank} rules out the case of redundant
dictionary elements by making sure its functions are linearly
independent. Moreover, it requires the data snapshots to be sampled in
a way that reflects this fact.

\begin{algorithm}
  \caption{Symmetric Subspace Decomposition} \label{algo:ssd}
  \begin{algorithmic}[1] 
    \Statex \textbf{Inputs:} $\dx, \dy \in \real^{N \times N_d }$ \quad \textbf{Output:} $\cssd$
    \State \textbf{Procedure} $\cssd \gets \ssd(\dx,\dy)$
    \State \textbf{Initialization} 
    \State $i \gets 1$, $A_1 \gets \dx$, $B_1 \gets \dy$, $\cssd \gets I_{N_d}$
    \While{1}
    \smallskip
    \State $\begin{bmatrix} Z_i^A \\ Z_i^B \end{bmatrix} \gets
    \operatorname{null}([A_i,B_i])$ \Comment{Basis for the null
      space}\smallskip \label{algossd:null}
    \If{$\operatorname{null}([A_i,B_i])=\emptyset$}
    \State \textbf{return} $0$ \Comment{The basis does not exist}
    \State \textbf{break}
    \EndIf
    \If{$\rown(Z_i^A)  \leq \coln(Z_i^A)$} \label{algossd:size-check}
    \State \textbf{return} $\cssd$ \Comment{The procedure is complete}
    \State \textbf{break}
    \EndIf
    \State $\cssd \gets \cssd Z_i^A$ \Comment{Reduce the subspace}
    \State $A_{i+1} \gets A_i Z_i^A$, $B_{i+1} \gets B_i Z_i^A$, $i \gets i+1$
    \EndWhile
  \end{algorithmic}
\end{algorithm}

The SSD algorithm provides an iterative approach to prune the
dictionary $D$ at each iteration by removing the functions that do not
correspond to linear evolutions.  The SSD algorithm terminates after
finite iterations and its output $\cssd$ satisfies the following
properties.
 
\begin{theorem}\longthmtitle{SSD
    Output~\citep[Theorem~5.1]{MH-JC:22-tac}}\label{t:ssd-convergence}
  Suppose that Assumption~\ref{a:full-rank} holds. Then,
  \begin{enumerate}
  \item $\cssd$ is either $0$ or has full column rank;
  \item $ \cssd$ satisfies $\range(\dx \cssd) = \range(\dy
    \cssd)$;
  \item the subspace $\range(\dx \cssd)$ is maximal, in the sense that,
    for any matrix $E$ with $\range(\dx E) = \range(\dy E)$, we have
    $\range(\dx E) \subseteq \range(\dx \cssd)$ and $\range(E) \subseteq
    \range(\cssd)$.
    \oprocend
  \end{enumerate}
\end{theorem}

The dictionary identified by SSD is
\begin{align}\label{eq:ssd-dictionary}
  \dssd(\cdot) := D(\cdot) \cssd .
\end{align}
Based on Theorem~\ref{t:ssd-convergence}, $\dssd$ spans the largest
subspace of $\Span(D)$ on which $\range(\dssd(X)) = \range(\dssd(Y))$.
One can apply the EDMD algorithm
(equations~\eqref{eq:EDMD-optimization}-\eqref{eq:EDMD-closed-form})
on $\ssddx$ and $\ssddy$ to find the predictor matrix
\begin{align}\label{eq:Kssd}
  \Kssd = \EDMD{\dssd}{X}{Y}= \ssddx^\dagger \ssddy.
\end{align}
The residual error $\| \ssddy - \ssddx \Kssd \|_F $ is equal to zero
and $\Kssd$ fully captures the behavior of the available
data. Moreover, one can replace $D$ and $\Kedmd$
in~\eqref{eq:function-predictor}-\eqref{eq:EDMD-eigenfunction} by
$\dssd$ and $\Kssd$ to define approximated Koopman eigenfunctions and
linear predictors for the dynamics.  Under reasonable assumptions on
the data sampling, $\Span(\dssd)$ is the maximal Koopman-invariant
subspace in $\Span(D)$ almost surely and consequently the
aforementioned eigenfunctions and predictors are \emph{exact} for
\emph{all points} (not just on the sampled data) in the state space
$\Mc$ almost surely. We refer the reader to~\citep{MH-JC:22-tac} for
additional information about the SSD algorithm, its convergence, and
its properties regarding the identification of the eigenfunctions of
the Koopman operator. We conclude this section by remarking that the
SSD algorithm is, in fact, an algebraic search.
  
\begin{remark}\longthmtitle{SSD is an Efficient Algebraic Subspace
    Search} {\rm One can view the SSD algorithm as an algebraic method
    that searches through all subspaces of a finite-dimensional vector
    space of functions. In fact, in Algorithm~\ref{algo:ssd}, one can
    view $\Span(D)$ as the space we search through, and the space
    $\Span(\dssd)$ (solution of SSD) as the maximal Koopman-invariant
    subspace (almost surely) in our search space $\Span(D)$. For
    example, if use the SSD algorithm on the system in
    Example~\ref{ex:more-not-better} and set $\Span(D_2)$ as our
    search space, we find the subspace $\Span(D_1)$, which is Koopman
    invariant and leads to exact prediction. \oprocend }
\end{remark}

\section{Problem Statement}\label{sec:problem-statement}
We start by noting that one can view the SSD and EDMD methods
described in Section~\ref{sec:preliminaries} as the two extreme cases
in the trade-off between prediction accuracy and dictionary
expressiveness. This is because, on the one hand, the SSD predictor
provides almost surely exact predictions for functions in
$\Span(\dssd)$ but, due to the pruning of the original dictionary~$D$,
this might not be sufficiently expressive to capture the system
dynamics.  EDMD, on the other hand, provides predictions for all
functions in $\Span(D)$, but there is no guarantee on the accuracy of
this prediction.

Our goal is then to explore the accuracy-expressiveness spectrum
in-between the extreme cases of SSD and EDMD.  To do this, we seek to
provide a formal data-driven characterization of how close a
function space is to being invariant under the Koopman operator
(something we refer to as \emph{invariance proximity}). Equipped with
this notion, we also aim to develop fast algebraic methods that can
find finite-dimensional function spaces that meet a desired level of
invariance proximity. Throughout the paper we do not assume any
information about the dynamical system except the existence of a
finite data set of snapshots gathered from its trajectories.

Formally, given the original dictionary $D$ defined
in~\eqref{eq:dictionary-def}, data snapshots matrices $X,Y$ gathered
from the dynamical system~\eqref{eq:dymamical-sys} defined
in~\eqref{eq:data-snapshots}, and assuming that
Assumption~\ref{a:full-rank} holds, we seek to:
\begin{enumerate}
\item provide a measure to quantify the invariance proximity of span
  of any dictionary $\tilde{D}$ with elements in $\Span(D)$ solely
  based on available data $X,Y$;
\item provide an algebraic algorithm that searches through the
  subspaces of $\Span(D)$ and finds a dictionary $\tilde{D}$, where
  $\Span(\tilde{D}) \subseteq \Span(D)$ meets a desired level of
  invariance proximity;
\item such that $\Span(\tilde{D})$ contains the maximal
  Koopman-invariant subspace in $\Span(D)$.
\end{enumerate}
Requirement~(c) ensures the correctness of the algorithmic solution by
ensuring the maximal Koopman-invariant subspace and all Koopman
eigenfunctions in $\Span(D)$ are captured.

\section{$\epsilon$-Apart Spaces Measure Invariance
  Proximity}\label{sec:epsilon-apart}
In this section we provide a quantifiable measure for invariance
proximity of a subspace by studying the behavior of EDMD with respect
to its dictionary. Since the true system dynamics is unknown, this
measure must be based on the available data matrices~$X$ and~$Y$.  To
gain a deeper understanding about the behavior of the data-dictionary
pair, we offer the following interpretation of the action of the
solution $\Kedmd$ of the optimization~\eqref{eq:EDMD-optimization} as
a projection from $\range(\dy)$ onto $\range(\dx)$. To see this, let
$w \in \range(\dy)$ be a vector of the form of $\dy v$.
Using~\eqref{eq:EDMD-closed-form},
\begin{align*}
  \dx \Kedmd v &= \dx \dx^\dagger \dy v
  \\
  &= \dx \dx^\dagger w = \Pc_{\dx} w,
\end{align*}
where we have used that $\dx \dx^\dagger$ is the projection operator
on $\range(\dx)$. Using this projection viewpoint
alongside~\eqref{eq:EDMD-optimization} reveals that the residual error
$\| \dy - \dx \Kedmd\|_F$ of EDMD, and consequently its accuracy on
the available data, depends of how close the subspaces $\range(\dx)$
and $\range(\dy)$ are. In fact, note that
\begin{itemize}
\item If $D$ spans a Koopman-invariant subspace, we have
  $\range(\dy)=\range(\dx)$ and the residual error of EDMD is equal to
  zero independently of the data (as long as
  Assumption~\ref{a:full-rank} holds). In this case, EDMD
    captures complete dynamical information about the evolution of all
    functions in $\Span(D)$ and the predictor
    in~\eqref{eq:function-predictor} is exact;
  \item instead, if $\range(\dx) \perp \range(\dy)$, one can deduce
    that under Assumption~\ref{a:full-rank},
    $\Kedmd = \mathbf{0_{N_d \times N_d}}$ and EDMD captures no
    information about the dynamics. In particular, the residual error
    $\| \dy - \Kedmd \dx\|_F = \|\dy\|_F$ amounts to $100 \%$
    prediction error for $\dy$.
\end{itemize}

We  illustrate next the aforementioned cases.
\begin{example}\longthmtitle{Dependence of EDMD's Prediction on Data and Dictionary}
  {\rm Consider the discrete-time system $x^+ = 1 -x$ with state space
    $\Mc = \real$. Suppose that we gather $2m$ data snapshots
    ($m \in \naturals$) from a single trajectory with length $2m +1$
    starting from the initial condition $x_0 = 0$. Hence,
    $X = [0,1,0,\ldots,1]^T$ and $Y = [1,0,1,\ldots,0]^T$. If we
    choose our dictionary as $D(x) = [1,x]$ (which spans a
    Koopman-invariant subspace), EDMD captures complete information
    about the dynamics. However, if we remove the constant function
    from the dictionary, the new dictionary snapshots matrices are
    equal to $X$ and $Y$, and span orthogonal subspaces. In this case,
    $\Kedmd = 0$ and EDMD does not capture any information about the
    dynamics.  \oprocend }
\end{example}

These observations suggests that the proximity of the vector spaces
$\range(\dx)$ and $\range(\dy)$ can be used as a quantifiable
characterization for invariance proximity of $\Span(D)$ and
consequently the prediction accuracy of EDMD. This motivates the
following definition.

\begin{definition}\longthmtitle{$\epsilon$-Apart
    Subspaces}\label{def:eps-apart}
  Given $\epsilon \geq 0$, two vector spaces
  $S_1,S_2 \subseteq \real^p$ are \emph{$\epsilon$-apart} if
  $ \| \Pc_{S_1}v - \Pc_{S_2}v \|_2 \leq \epsilon \|v\|_2$, for all
  ${v \in S_1 \cup S_2}$.  \oprocend
\end{definition}

According to this definition\footnote{
Note that, unlike Grassmannians,
  e.g.~\citep{PAA-RM-RS:09}, there is no restriction on the dimension
  of the subspaces in Definition~\ref{def:eps-apart}.}, the norm of
the error induced by projecting a vector $v$ belonging to one of the
subspaces onto the other subspace is smaller than $\epsilon \|v\|_2$.  Next, we show that this
notion fully characterizes equality of spaces with the case $\epsilon
=0$.

\begin{lemma}\longthmtitle{0-apart Subspaces are
    Equal}\label{l:zero-apart-equal}
  Vector spaces $S_1,S_2 \subseteq \real^p$ are 0-apart if and only if
  $S_1 =S_2$.
\end{lemma}
\begin{pf}
  $(\Rightarrow)$: Let $v \in S_1$.  By definition, $ \| \Pc_{S_1}v -
  \Pc_{S_2}v \|_2 = \| v - \Pc_{S_2}v \|_2 =0$, and hence $ v =
  \Pc_{S_2}v$.  Consequently, $v \in S_2$, showing $S_1 \subseteq
  S_2$. The inclusion $S_2 \subseteq S_1$ can be proved analogously,
  and we conclude $S_1 = S_2$.
  
  $(\Leftarrow)$: Since $S_1 = S_2$, for all $v \in S_1 = S_2$, we
  have $\Pc_{S_1}v = \Pc_{S_2}v = v$. Hence, $ \| \Pc_{S_1}v -
  \Pc_{S_2}v \|_2 =0$, for all $v \in S_1 \cup S_2$, and the result
  follows. \qed
\end{pf}

The next result shows that all subspaces are $1$-apart.

\begin{lemma}\longthmtitle{Any Two Subspaces are
    1-apart}\label{l:all-subs-one-apart}
  Any two vector spaces $S_1,S_2 \subseteq \real^p$ are 1-apart.
\end{lemma}
\begin{pf}
  For any $v \in S_1$, one can write
  \begin{align*}
    \| \Pc_{S_1}v - \Pc_{S_2}v \|_2 = \| v - \Pc_{S_2}v \|_2 = 
    \| (I-\Pc_{S_2})v \|_2 \leq \|v\|_2,
  \end{align*}
  where in the last equality we have used the fact that
  $(I-\Pc_{S_2})$ is the projection operator on the orthogonal
  complement of~$S_2$.  One can write a similar argument for $v \in
  S_2$, which completes the proof. \qed
\end{pf} 

Lemmas~\ref{l:zero-apart-equal}-\ref{l:all-subs-one-apart} together
imply that the range $ [0,1]$ for the parameter~$\epsilon$ fully
characterizes the proximity of any two subspaces. This enables us to
use the concept of $\epsilon$-apart subspaces on $\dx$ and $\dy$ as a
way to quantify the invariance proximity of $\Span(D)$ under the
Koopman operator associated with the system~\eqref{eq:dymamical-sys}.
Equipped with this, we reformulate next the problems (b)-(c) in
Section~\ref{sec:problem-statement}.

\begin{problem}\longthmtitle{Balancing Prediction Accuracy and
    expressiveness}\label{prob:eps-apart-identification}
  Given the parameter $\epsilon \in [0,1]$, find a dictionary
  $\tilde{D}$ with elements in $\Span(D)$ such that
  \begin{enumerate}
  \item[(b)] $\range(\tdx)$ and $\range(\tdy)$ are $\epsilon$-apart;
  \item[(c)] $\Span(\tilde{D})$ contains the maximal Koopman-invariant
    subspace in $\Span(D)$.  \oprocend
  \end{enumerate}
\end{problem} 

It is worth mentioning that 
\begin{align*}
  \epsilon^*= \min \setdef{\epsilon \in [0,1]}{\range(\dx), \range(\dy) \text{ are
      $\epsilon$-apart}}
\end{align*}
captures the invariance proximity, and consequently the prediction
accuracy, of~$D$. As a result, if we choose $\epsilon < \epsilon^*$ in
Problem~\ref{prob:eps-apart-identification}, the new dictionary would
be smaller than $D$, leading to a decrease of the expressiveness of
the resulting dictionary. Hence, the choice of parameter~$\epsilon$
strikes a balance between prediction accuracy and expressiveness of
the dictionary.

\section{Tunable Symmetric Subspace Decomposition}\label{sec:TSSD}
In this section, we design and analyze an algorithm, termed Tunable
Symmetric Subspace Decomposition (T-SSD), to address
Problem~\ref{prob:eps-apart-identification}.

\subsection{The T-SSD Algorithm}
Given the dictionary $D$ and data snapshots $X$, $Y$, the problem of
finding a dictionary $\tilde{D}$ such that $\range(\tdx)$ and
$\range(\tdy)$ are $\epsilon$-apart can be tackled by pruning~$D$.  We
next describe informally the procedure and then formalize it in
Algorithm~\ref{algo:tssd}.

[\emph{Informal description}:] The pruning consists of
  identifying the functions that violate the desired invariance
  proximity condition and remove them from the span of the dictionary. To identify
  such functions, we define the projection difference matrix
  (Step~\ref{algotssd:G} in Algorithm~\ref{algo:tssd})
  \begin{align*}
    G = \Pc_{\dx}- \Pc_{\dy} = \dx \dx^\dagger - \dy \dy^\dagger,
  \end{align*}
  which is a symmetric matrix with mutually orthogonal eigenvectors
  spanning $\real^N$ (with corresponding real-valued
  eigenvalues). Interestingly, if all eigenvalues of $G$ belong to
  $[-\epsilon, \epsilon]$, then $\dx$ and $\dy$ are
  $\epsilon$-apart. Otherwise, we focus our attention on the smaller
  subspace of $\real^N$ defined by
  \begin{align*}
    \Wc_\epsilon := \Span \setdef{v \in \real^N}{G v = \lambda v, \,
      |\lambda| \leq \epsilon},
  \end{align*}
  corresponding to the span of eigenvectors of $G$ with eigenvalues in
  $[-\epsilon, \epsilon]$.  For practical reasons, we work with a
  basis for $\Wc_\epsilon$ (Step~\ref{algotssd:Vi} in
  Algorithm~\ref{algo:tssd}). Next, we find the largest dictionary
  $\tilde{D}$ with elements in $\Span(D)$ such that $\range(\tdx),
  \range(\tdy) \subseteq \Wc_\epsilon$
  (Steps~\ref{algotssd:symmetric-intersection}-\ref{algotssd:sub-reduction}
  in Algorithm~\ref{algo:tssd}). There are two possible outcomes:
  \begin{enumerate}[label=(\roman*)]
  \item $\dim \tilde{D} = \dim D$;
  \item $\dim \tilde{D} < \dim D$.
  \end{enumerate}
  Scenario (i) indicates that the dictionary $D$ does not need pruning
  and $\range(\dx), \range(\dy)$ are $\epsilon$-apart
  (Steps~\ref{algotssd:size-check}-\ref{algotssd:break-complete} in
  Algorithm~\ref{algo:tssd}). On the other hand, scenario~(ii) leads
  to a dictionary of lower dimension.
  However, it is not guaranteed that $\range(\tdx)$ and $\range(\tdy)$
  are $\epsilon$-apart since $\tilde{D}$ is a different dictionary
  than~$D$. Consequently, we re-run the process, starting with the
  definition of $G$, for the new dictionary $\tilde{D}$. This leads to
  an iterative implementation that stops when the dictionary cannot be
  reduced anymore (yielding the desired $\epsilon$-apart subspaces). 

  The formalization of this procedure yields the Tunable Symmetric
  Subspace Decomposition (T-SSD)\footnote{In
    Algorithms~\ref{algo:tssd}-\ref{algo:symmetric-intersection}, the
    outputs of $\operatorname{null}(A)$ and $\basis(A)$ are matrices
    whose columns form orthonormal bases for the null space of $A$ and
    $\range(A)$, respectively.} in Algorithm~\ref{algo:tssd}.  We make
  the following additional observations regarding the use of notation
  to provide intuition about the algorithm pseudocode: (i) we index
  the internal matrix variables based on the iteration number (this
  facilitates later the in-depth algebraic analysis); (ii)
noting that, at each iteration, the dictionary elements are linear
combinations of the elements of the original dictionary, we represent
the dictionary at iteration $i$ simply by a matrix $C_i$, which
corresponds to the dictionary $D(\cdot) C_i$; (iii) using the
representation in (ii), we do not need to form the dictionary and
apply it on the data matrices~$X$ and~$Y$. Instead, the effect of the
dictionary at iteration $i$ on the data can be represented as
$A_i = \dx C_i$ and $B_i= \dy C_i$.

\begin{algorithm}[thb]
  \caption{Tunable Symmetric Subspace Decomposition}\label{algo:tssd}
  \begin{algorithmic}[1]
    \Statex \textbf{Inputs:} $\dx, \dy \in \real^{N \times N_d }$, $\epsilon \in [0,1]$
    \vspace{5pt}
    \State \textbf{Procedure} $\tssd(\dx,\dy,\epsilon)$
    \State \textbf{Initialization} 
    \State $i \gets 0$, $A_0 \gets \dx$, $B_0 \gets \dy$, $C_0\gets I_{N_d}$
    \While{1}
    \State $i \gets i+1$
    \State $G_{i} \gets A_{i-1} A_{i-1}^\dagger - B_{i-1} B_{i-1}^\dagger$  \label{algotssd:G}
    \Statex  \hspace{14pt}$\triangleright$ projection difference \smallskip
    \State $V_i \gets \basis(\Span\setdef{v \in \real^N}{G_{i} v = \lambda v, \, |\lambda| \leq \epsilon})$ \label{algotssd:Vi}
    \Statex  \hspace{14pt}$\triangleright$ eigenpairs corresponding to small eigenvalues\smallskip
    \State $E_i \gets \symmintersection(V_i,A_{i-1},B_{i-1})$\label{algotssd:symmetric-intersection}
    \Statex  \hspace{14pt}$\triangleright$ Find largest dictionary matrices in $V_i$ (Algorithm~\ref{algo:symmetric-intersection}) \smallskip
    \State $C_i \gets C_{i-1} E_i$ \Comment{reduce  subspace} \label{algotssd:sub-reduction}
    \State $A_{i} \gets A_{i-1} E_i$, $B_{i} \gets B_{i-1} E_i$
    \Statex  \hspace{14pt}$\triangleright$ calculate  new dictionary matrices\smallskip
    
    \If{$E_i = 0$} \label{algotssd:zero-check}
    \State \textbf{return} $0$ \label{algotssd:sub-not-exist}
    \Statex  \hspace{14pt}$\triangleright$ subspace does not exist, returning scalar $0$ \smallskip
    \State \textbf{break} \label{algotssd:break-zero}
    \EndIf
    
    \If{$\rown(E_i) \leq \coln(E_i)$} \label{algotssd:size-check}
    \State \textbf{return} $C_i$ \Comment{procedure is complete} \label{algotssd:complete}
    \State \textbf{break} \label{algotssd:break-complete}
    \EndIf
    \EndWhile
  \end{algorithmic}
\end{algorithm}

Algorithm~\ref{algo:symmetric-intersection} describes the
$\symmintersection$ function in
Step~\ref{algotssd:symmetric-intersection} of T-SSD: this strategy
corresponds to the computation described above of the largest
dictionary $\tilde{D}$ such that $\range(\tdx)$ and $\range (\tdy)$
belong to the reduced subspace~$W_\epsilon$. Similarly to
Algorithm~\ref{algo:tssd}, instead of actually forming the reduced
dictionary, Algorithm~\ref{algo:symmetric-intersection} uses the
matrix-based representation of the dictionary.  Next, we explain the
steps of the algorithm and the reason behind its naming. Given input
matrices $V$, $A$, and $B$, Step~\ref{algosi:first-null} in
Algorithm~\ref{algo:symmetric-intersection} identifies $W_A$ such that
$\range(AW_A) = \range(V) \cap \range(A)$ (see
Lemma~\ref{l:subspace-intersection}).  Then, again in
Step~\ref{algosi:return-basis} the algorithm (by virtue of
Lemma~\ref{l:subspace-intersection}) finds the matrix $Z_B$ such that
$\range(B W_A Z_B) = \range(V) \cap \range(B W_A)$. The output matrix
$E:=\basis(W_A Z_B)$ (cf.~Step~\ref{algosi:return-basis}) then
specifies the largest subspaces $\range(AE)$, $\range(BE)$ both
belonging to $\range(V)$.  Note the symmetry in this specification: if
a linear combination of the columns of $A$ is in $\range (V)$, then
the same linear combination of columns of $B$ belongs to $\range(V)$.
Moreover, Algorithm~\ref{algo:symmetric-intersection} breaks and
returns $0$ if any of the aforementioned intersections only contain
the zero vector (Steps~\ref{algosi:first-if}-\ref{algosi:first-break}
and Steps~\ref{algosi:second-if}-\ref{algosi:second-break}).

\begin{algorithm}[thb]
  \caption{Symmetric Intersection} \label{algo:symmetric-intersection}
  \begin{algorithmic}[1]
    \Statex \textbf{Inputs:} $V \in \real^{n \times m }$ and $A,B \in \real^{n \times p}$
    \vspace{5pt}
    \State \textbf{Procedure} $\symmintersection(V,A,B)$
    \If{$\operatorname{null}([V,A])=\emptyset$} \label{algosi:first-if}
    \State \textbf{return} $0$
    \State \textbf{break}  \label{algosi:first-break}
    \Else 
    \smallskip
    \State $\begin{bmatrix} W_V \\ W_A \end{bmatrix} \gets
    \operatorname{null}([V,A])$ \label{algosi:first-null} 
    \smallskip
    \Statex  \hspace{12pt}$\triangleright$ $\coln(V)=\rown(W_V), \coln(A) = \rown(W_A)$
    \If{$\operatorname{null}([V,BW_A])=\emptyset$} \label{algosi:second-if}
    \State \textbf{return} 0
    \State \textbf{break} \label{algosi:second-break}
    \EndIf
    \smallskip
    \State $\begin{bmatrix} Z_V \\ Z_B \end{bmatrix} \gets
    \operatorname{null}([V,B W_A])$ \label{algosi:second-null} 
    \smallskip
    \Statex  \hspace{12pt}$\triangleright$ $\coln(V)=\rown(Z_V),
    \coln(B W_A) = \rown(Z_B)$ 
    \EndIf
    \State \textbf{return} $\basis(W_A Z_B)$  \Comment{Returning an
      orthogonal basis}  \label{algosi:return-basis} 
  \end{algorithmic}
\end{algorithm}

\begin{remark}\longthmtitle{Implementation of
    Algorithm~\ref{algo:symmetric-intersection} on Finite-Precision
    Computers}
  {\rm The accuracy of the implementation of
    Algorithm~\ref{algo:symmetric-intersection} depends on the
    calculation of the null space of several matrices, which might be
    sensitive to round-off errors. To circumvent this issue, one can
    set sufficiently small (according to a desired accuracy level)
    singular values of the matrices to zero.
    \oprocend }
\end{remark}

\subsection{Basic Properties of T-SSD}

Our end goal now is to show that the T-SSD algorithm solves
Problem~\ref{prob:eps-apart-identification} and unveil its
relationship with the EDMD and SSD methods. In order to do so, we
establish here several basic algorithm properties.

\begin{proposition}\longthmtitle{Properties of
    $\symmintersection$}\label{p:symmetric-intersection}
  Let matrices $V,A,B$ have full column rank and $E =
  \symmintersection(V,A,B)$. Then,
  \begin{enumerate}
  \item $E = 0$ or $E^T E = I$;
  \item $ \range(AE),\range(BE) \subseteq \range(V)$;
  \item $E$ is maximal, i.e., any nonzero matrix $F$ such that $
    \range(AF) ,\range(BF) \subseteq \range(V)$ satisfies $\range(F)
    \subseteq \range(E)$.
  \end{enumerate}
\end{proposition}
\begin{pf}
  (a) There are three ways for
  Algorithm~\ref{algo:symmetric-intersection} to terminate. If the
  algorithm executes
  Steps~\ref{algosi:first-if}-\ref{algosi:first-break} or
  Steps~\ref{algosi:second-if}-\ref{algosi:second-break}, we have $E
  =0$ by definition. Otherwise, the algorithm executes
  Step~\ref{algosi:return-basis}. Hence, noting that $W_A$ and $Z_B$
  exist and the basis function returns an orthonormal basis for $W_A
  Z_B$, one can conclude $E^T E= I$.
	
  (b) The case $E = 0$ is trivial. Suppose that $E \neq 0 $ and hence
  has full column rank according to part~(a). By definition,
  $\range(E) = \range(W_A Z_B)$. Consequently, based on
  Step~\ref{algosi:second-null} of the algorithm and using
  Lemma~\ref{l:subspace-intersection}, we deduce
  \begin{align}\label{eq:algosi-range-inequality-1}
    \range(B E) &= \range(BW_A Z_B)
    \nonumber \\
    &= \range(BW_A ) \cap \range(V) \subseteq \range(V),
  \end{align}
  where in the first equality, we used
  Lemma~\ref{l:product-subspace}. Moreover, from the definition
  of~$E$, one can deduce that $\range(E) \subseteq \range(W_A)$. In
  addition, based on Lemma~\ref{l:product-subspace}, we have
  $\range(AE) \subseteq \range(A W_A)$. Using the previous inequality
  in conjunction with Lemma~\ref{l:subspace-intersection} applied to
  Step~\ref{algosi:first-null} of the algorithm, one can write
  \begin{align}\label{eq:algosi-range-inequality-2}
    \range(AE) \subseteq \range(A W_A) = \range(A) \cap \range(V)
    \subseteq \range(V),
  \end{align}
  which in conjunction with~\eqref{eq:algosi-range-inequality-1}
  concludes the proof of~(b).
  
  (c) Without loss of generality, we assume that $F$ has full column
  rank (if that is not the case, one can consider another matrix
  $\bar{F}$ with full column rank such that $\range(F) =
  \range(\bar{F})$). Since $\range(AF) \subseteq \range(V)$, we have
  $\range(AF) \subseteq \range(A) \cap \range(V)$, which leads to
  $\range(AF) \subseteq \range(A W_A)$ based
  on~\eqref{eq:algosi-range-inequality-2}. Moreover, one can use
  Lemma~\ref{l:product-subspace} to deduce that $\range(F) \subseteq
  \range(W_A)$. Since $F$ and $W_A$ both have full column rank, there
  exists $F_W$ with full column rank such that
  \begin{align}\label{eq:algosi-F-decomposition}
    F = W_A F_W.
  \end{align}
  Considering that $\range(BF) \subseteq \range(V)$ and $\range(B W_A
  F_W) \subseteq \range(B W_A )$ in combination
  with~\eqref{eq:algosi-F-decomposition}, we deduce $ \range(BF)
  =\range(B W_A F_W) \subseteq \range(B W_A) \cap \range(V) = \range(B
  W_A Z_B)$, where the last equality follows
  from~\eqref{eq:algosi-range-inequality-1}. Based on
  Lemma~\ref{l:product-subspace}, we deduce $\range(F) \subseteq
  \range(W_A Z_B) = \range(E)$. \qed
\end{pf}

Next, we show that T-SSD terminates after a finite number of
iterations.

\begin{proposition}\longthmtitle{Finite-time Termination of T-SSD
    Algorithm}\label{p:finite-time-termination}
  The T-SSD algorithm terminates after at most $N_d$ iterations.
\end{proposition}
\begin{pf}
  We reason by contradiction. Suppose that the algorithm does not
  terminate before iteration $N_d +1$. Hence, the algorithm does not
  execute Steps~\ref{algotssd:sub-not-exist}-\ref{algotssd:break-zero}
  or Steps~\ref{algotssd:complete}-\ref{algotssd:break-complete} in
  the first $N_d$ iterations. Therefore, the conditions in
  Steps~\ref{algotssd:zero-check} and~\ref{algotssd:size-check} do not
  hold. Using Proposition~\ref{p:symmetric-intersection}(a), one can
  write
  \begin{align}\label{eq:E-dim-reduction}
    \rown(E_i) > \rown(E_i)-1 \geq \coln(E_i), 
  \end{align}
  for all $ i \in \until{N_d}$.  In addition, based on the definition
  of the $E_i$'s, one can deduce $ \coln(E_i) = \rown(E_{i+1})$, for
  all $i \in \until{N_d}$.  Combining this
  with~\eqref{eq:E-dim-reduction} leads to $ \rown(E_1) \geq \coln(E_{N_d}) +
  N_d$.  This fact together with $\rown(E_1)= N_d$ and $
  \coln(E_{N_d}) = \coln(C_{N_d})$
  (cf. Step~\ref{algotssd:sub-reduction}) implies that $\coln(C_{N_d})
  \leq 0$, contradicting $\coln(C_{N_d})\ge 1$. \qed
\end{pf}

Next, we study basic properties of the internal matrices of the T-SSD
algorithm.

\begin{lemma}\longthmtitle{Properties of T-SSD
    Matrices}\label{l:tssd-basic-properties}
  Let the T-SSD algorithm terminate in $L$ time steps. Then,
  \begin{enumerate}
  \item $\forall i \in \{0,\ldots, L-1\},\; \range(C_{i+1}) \subseteq
    \range(C_i)$;
  \item $\forall i \in \{0,\ldots, L-1\},\; C_i^T C_i = I$;
  \item $C_L = \mathbf{0}$ or $C_L^T C_L =I$,
  \end{enumerate}
  where $C_i$ denotes T-SSD's $i$th internal matrix, cf.
    Algorithm~\ref{algo:tssd}.
\end{lemma}
\begin{pf}
  (a) According to Step~\ref{algotssd:sub-reduction} of the algorithm,
  $C_{i+1} = C_i E_{i+1}$. Hence, $\range(C_{i+1}) =
  \range(C_{i}E_{i+1}) \subseteq \range(C_i)$.
  
  (b) For $i=0$, the result holds by definition. Moreover, since the
  algorithm does not terminate until iteration $L$, it does not
  execute Steps~\ref{algotssd:sub-not-exist}-\ref{algotssd:break-zero}
  in iterations $\{1,\ldots,L-1\}$. Hence, $E_i \neq 0$ and based on
  Proposition~\ref{p:symmetric-intersection}(a), we have
  \begin{align}\label{eq:E-orthogonal}
    E_i^T E_i = I, \, \forall i \in \until{L-1}.
  \end{align}
  Moreover, from Step~\ref{algotssd:sub-reduction}, $ C_i = C_0 E_1
  E_2 \cdots E_i, \, \forall i \in \until{L-1}$.  This in conjunction
  with~\eqref{eq:E-orthogonal} and $C_0 = I_{N_d}$, implies $C_i^T C_i
  = I$ for all $i \in \until{L-1}$, as claimed.
  
  (c) Note that $C_L =C_{L-1} E_L$. Based on
  Proposition~\ref{p:symmetric-intersection}(a), either $E_L = 0$ or
  $E_L^T E_L =I$. In the former case, we have $C_L = \mathbf{0}$. In
  the latter case, $ C_L^T C_L = C_{L-1}^T E_L^T E_L C_{L-1} =
  C_{L-1}^T C_{L-1} = I$, where in the last equality we used~(b). \qed
\end{pf}

For convenience, let
\begin{align}\label{eq:ctssd}
  \ctssd := \tssd(\dx,\dy,\epsilon),
\end{align}
denote the output of the T-SSD algorithm. This leads to the definition
of the T-SSD dictionary
\begin{align}\label{eq:reduced-dictionary}
  \dtssd(\cdot) := D(\cdot) \ctssd.
\end{align}
To extract the dynamical information associated with the Koopman
operator on $\Span(\dtssd)$, we use EDMD.  According
to~\eqref{eq:EDMD-closed-form}, we find the T-SSD prediction
  matrix as
\begin{align}\label{eq:Ktssd}
  \Ktssd := \EDMD{\dtssd}{X}{Y} = \dtssdx ^ \dagger \dtssdy.
\end{align}
We can also define approximated Koopman eigenfunctions according
to~\eqref{eq:EDMD-eigenfunction} using the eigendecomposition of
$\Ktssd$ and the dictionary~$\dtssd$.
In addition, following~\eqref{eq:function-predictor}, given any
  function $f \in \Span(\dtssd)$ described by
  $f(\cdot) = \dtssd(\cdot) w$, we can define the T-SSD predictor of
  $\Kc f$ on $\Span(\dtssd)$ as
\begin{align}\label{eq:predictor-tssd}
  \Pf_{\Kc f}^{\tssd}(\cdot) = \dtssd(\cdot) \Ktssd w.
\end{align}

\begin{remark}\longthmtitle{Computational Complexity of
    T-SSD}\label{r:tssd-complexity}
  {\rm Given $N$ data snapshots and $N_d$ dictionary functions, and
    considering the complexity of scalar operations as $O(1)$, the
    most time-consuming step of Algorithm~\ref{algo:tssd} is
    Step~\ref{algotssd:Vi}, which requires the eigendecomposition of
    an $N \times N$ matrix and takes $O(N^3)$ operations. Based on
    Proposition~\ref{p:finite-time-termination}, the algorithm
    terminates after at most $N_d$ iterations, resulting in a total
    complexity of $O(N^3 N_d)$.  \oprocend }
\end{remark}

\section{T-SSD Balances Accuracy and Expressiveness}
In this section we show that the output of T-SSD balances prediction
accuracy and expressiveness as prescribed by the design
parameter~$\epsilon \in [0,1]$.

\subsection{T-SSD Identifies $\epsilon$-Apart Subspaces}
Here, we show that T-SSD solves
Problem~\ref{prob:eps-apart-identification}(b).

\begin{theorem}\longthmtitle{T-SSD Output Subspaces are
    $\epsilon$-Apart}\label{t:tssd-output-eps-apart}
  $\range(\dtssdx) $ and $\range(\dtssdy)$ are $\epsilon$-apart.
\end{theorem}
\begin{pf}
  Let $L \leq N_d$ be the number of iterations for convergence of the
  T-SSD algorithm
  (cf. Proposition~\ref{p:finite-time-termination}). Based on
  Proposition~\ref{p:symmetric-intersection}(a), we have $E_L=0$ or
  $E_L^T E_L = I$. With the notation of
    Algorithm~\ref{algo:tssd}, in the former case, the algorithm
  executes
  Steps~\ref{algotssd:sub-not-exist}-\ref{algotssd:break-zero} at
  iteration $L$ and consequently $\ctssd = 0$. Therefore,
  \begin{align*}
    \range(\dtssdx) = \range(\dtssdy) =\{\mathbf{0}_{N\times 1} \},
  \end{align*}
  and the result holds trivially.  Now, suppose that $E_L^T E_L =
  I$. Hence, $E_L$ has full column rank and consequently $\rown(E_L)
  \geq \coln(E_L)$.  However, since the algorithm executes
  Steps~\ref{algotssd:complete}-\ref{algotssd:break-complete}, the
  condition in Step~\ref{algotssd:size-check} holds and one can write
  $ \rown(E_L) = \coln(E_L)$. Therefore, since $E_L$ has full column
  rank, it is a nonsingular square matrix and
  \begin{subequations}\label{eq:range-equality-last-two-steps}
    \begin{align}
      \range(C_L) = \range(C_{L-1} E_L)= \range(C_{L-1}),
      \\
      \range(A_{L}) = \range(A_{L-1} E_L) = \range(A_{L-1}),
      \\
      \range(B_{L}) = \range(B_{L-1} E_L) = \range(B_{L-1}).
    \end{align}
  \end{subequations}
  At iteration $L$, one can use Steps~\ref{algotssd:G}
  and~\ref{algotssd:Vi} 
  in conjunction with the fact that the eigenvectors of $G_L$ are
  mutually orthogonal to write
  \begin{align}\label{eq:last-iteration-proj-diff}
    \|G_{L} v \|_2 &= \| A_{L-1} A_{L-1}^\dagger v - B_{L-1}
    B_{L-1}^\dagger v \|_2 \nonumber
    \\
    &= \| \Pc_{A_{L-1}} v - \Pc_{B_{L-1}} v \|_2 \leq \epsilon
    \|v\|_2,
  \end{align}
  for all $v \in \range(V_L)$.  Moreover, based on definition of $E_L$
  and Proposition~\ref{p:symmetric-intersection}(b),
  \begin{align}\label{eq:range-ET-Vi}
    \range(A_{L-1}E_L), \range(B_{L-1}E_L) \subseteq \range(V_L).
  \end{align}
  Consequently, using $A_L = \dx C_L$ and $B_L = \dy C_L$, and
  equations~\eqref{eq:range-equality-last-two-steps}-\eqref{eq:range-ET-Vi},
  we deduce
  \begin{align*}
    \| \Pc_{\dx C_L} v - \Pc_{\dy C_L} v \|_2 &= \| \Pc_{A_L} v -
    \Pc_{B_L} v \|_2
    \\
    = \| \Pc_{A_{L-1}} v - \Pc_{B_{L-1}} v \|_2 &\leq \epsilon
    \|v\|_2,
  \end{align*}
  for all $v \in \range(\dx C_L) \cup \range(\dy C_L)$.  Since $C_L =
  \ctssd$, and given the definition~\eqref{eq:reduced-dictionary} of
  the T-SSD dictionary, this can be rewritten as $ \| \Pc_{\dtssdx}
  v - \Pc_{\dtssdy} v \|_2 \leq \epsilon \|v\|_2$, for all $v \in
  \range(\dtssdx) \cup \range(\dtssdy)$. Hence, $\range(\dtssdx)$ and
  $\range(\dtssdy)$ are $\epsilon$-apart. \qed
\end{pf}

We next build on Theorem~\ref{t:tssd-output-eps-apart} to characterize
the accuracy of predictions~\eqref{eq:predictor-tssd} for any function
in $\Span(\dtssd)$ on the available data.

\begin{theorem}\longthmtitle{Relative Root Mean Square Error (RRMSE) of 
    Koopman Predictions by T-SSD are Bounded by $\epsilon$}\label{t:rrms-bound}
  For any function $f \in \Span(\dtssd)$, 
  \begin{align}\label{eq:relative-rms-error}
    \frac{\sqrt {\frac{1}{N} \sum_{i=1}^N | \Kc f (x_i) - \Pf_{\Kc f}^{\tssd}(x_i)|^2 } }{ \sqrt {\frac{1}{N} \sum_{i=1}^N |\Kc f (x_i)|^2 } } \leq \epsilon 
  \end{align}
  where $x_i^T$ is the $i$th row of $X$ and $\Pf_{\Kc f}^{\tssd}$ is defined
  in~\eqref{eq:predictor-tssd}.
\end{theorem}
\begin{pf}
  For convenience, we use the compact notation $\tilde{D}$ to refer to
  $\dtssd$ throughout the proof. We first prove the statement for
  real-valued functions in $\Span (\tilde{D})$. Let $f (\cdot) =
  \tilde{D}(\cdot) w$ with $w \in \real^{\coln(\ctssd)}$. From
  Theorem~\ref{t:tssd-output-eps-apart}, one can write $ \| (
  \Pc_{\tdy} - \Pc_{\tdx} ) v \|_2 \leq \epsilon \|v\|_2$, for all $v
  \in \range(\tdx) \cup \range(\tdy)$.  
  One can rewrite this equation as
  \begin{align}\label{eq:error-bound-generic}
    \| ( \tdy \tdy^\dagger - \tdx \tdx^\dagger ) v \|_2 \leq \epsilon \|v\|_2,
  \end{align}
  for all $v \in \range(\tdx) \cup \range(\tdy)$.  In addition, using
  equations~\eqref{eq:Ktssd} and~\eqref{eq:predictor-tssd},
  and the fact that $\Kc f(x_i) =f\circ T(x_i) = f(y_i)$ for all $i \in \until{N}$,
  one can write
  \begin{align*}
    & \sqrt {\sum_{i=1}^N \big| \Kc f  (x_i) - \Pf_{\Kc f}^{\tssd}(x_i)\big|^2 } = \|
    (\tdy - \tdx \Ktssd) w \|_2
    \\
    & \qquad = \| (\tdy - \tdx \tdx^\dagger \tdy) w \|_2
    \\
    & \qquad = \| \big( \tdy \tdy^\dagger - \tdx \tdx^\dagger \big)
    \tdy w\|_2 ,
  \end{align*}
  where we have used $\tdy = \tdy \tdy^\dagger \tdy$ in the last
  equality.  Moreover, since
  $\tdy w \in \range(\tdx) \cup \range(\tdy)$, one can use this
  equation in conjunction with~\eqref{eq:error-bound-generic} to write
  \begin{align*}
    \sqrt {\sum_{i=1}^N \big| \Kc f  (x_i) - \Pf_{\Kc f}^{\tssd}(x_i) \big|^2 } &\leq
    \epsilon \|\tdy w\|_2
    \nonumber \\
    & = \epsilon \sqrt { \sum_{i=1}^N | \Kc f (x_i)|^2 } .
  \end{align*}
  Scaling both sides by $N^{-\frac{1}{2}}$
  yields~\eqref{eq:relative-rms-error} for real-valued functions in
  $\Span(\tilde{D})$.
	
  For the complex-valued case, let $f(\cdot) = \tilde{D}(\cdot)w$ with
  $w = w_{\re} +j w_{\im}$, $w_\re, w_\im \in \real^{\coln(\ctssd)}$
  and $w_\im \neq 0$. 
  
 Consider the decompositions of $f$ and $\Pf_{\Kc f}^{\tssd}$ as
    $f(\cdot) = f_{\re}(\cdot) + j f_{\im}(\cdot)$ and
    $\Pf_{\Kc f}^{\tssd}(\cdot) = \Pf_{\Kc f_{\re}}^{\tssd}(\cdot) + j
    \Pf_{\Kc f_{\im}}^{\tssd}(\cdot)$, where
    \begin{align}\label{eq:function-decomposition-2}
      &f_\re(\cdot) = \tilde{D}(\cdot) w_\re, && f_\im(\cdot) =
                                                 \tilde{D}(\cdot) w_\im,
      \\
      &\Pf_{\Kc f_{\re}}^{\tssd}(\cdot) = \tilde{D}(\cdot) \Ktssd
        w_\re, &&\Pf_{\Kc f_{\im}}^{\tssd}(\cdot) = \tilde{D}(\cdot)
                  \Ktssd w_\im. \nonumber
  \end{align}
  Using~\eqref{eq:relative-rms-error} for the real-valued functions
  in~\eqref{eq:function-decomposition-2},
  \begin{align*}
    \sum_{i =1}^{N} |\Kc f_{\re} (x_i)-\Pf_{\Kc f_{\re}}^{\tssd}(x_i) |^2 \leq
    \epsilon^2 \sum_{i =1}^{N} |\Kc f_{\re} (x_i)|^2,
    \\
    \sum_{i =1}^{N} |\Kc f_{\im} (x_i)- \Pf_{\Kc f_{\im}}^{\tssd}(x_i) |^2 \leq
    \epsilon^2 \sum_{i =1}^{N} |\Kc f_{\im}(x_i)|^2.
  \end{align*}
  By adding these two inequalities,
  using~\eqref{eq:function-decomposition-2}, and noting that
  $|g|^2 = |g_\re|^2 + |g_\im|^2$ for $g = g_\re + j g_\im$, one can
  write
  \begin{align*}
    \sum_{i =1}^{N} |\Kc f(x_i)- \Pf_{\Kc f}^{\tssd}(x_i) |^2 \leq
    \epsilon^2 \sum_{i =1}^{N} |\Kc f (x_i)|^2,
  \end{align*}
  and~\eqref{eq:relative-rms-error} follows.  \qed
\end{pf}

Theorem~\ref{t:rrms-bound} ensures that each member of the vector
space of functions identified by T-SSD has prediction error bounded by
the accuracy parameter~$\epsilon$.
  
\begin{remark}\longthmtitle{T-SSD Bounds the Relative $L_2$-norm Error
    of Koopman Predictions under Empirical Measure by $\epsilon$}
  {\rm Given the functions in $\Span(D)$ and their composition with
    $T$ are measurable, consider the empirical measure
    $\mu = \frac{1}{N}\sum_{k=1}^{N} \delta_{x_k}$, where
    $\delta_{x_k}$ denotes the Dirac delta function at $x_k$, the
    $k$th row of~$X$. Then Theorem~\ref{t:rrms-bound} can be
    rewritten as
    \begin{align*}
      \frac{\|\Kc f - \Pf_{\Kc f}^{\tssd}\|_{L_2}}{\|\Kc f\|_{L_2}}
      \leq \epsilon, \; \forall f \in \Span(\dtssd), 
    \end{align*}
    where the $L_2$-norm is calculated with respect to the empirical
    measure $\mu$.}  \oprocend
\end{remark}

\subsection{T-SSD Captures Maximal Koopman-Invariant Subspace}

Here, we show that T-SSD also solves
Problem~\ref{prob:eps-apart-identification}(c). To do this, we study
the relationship of the algorithm with Koopman eigenfunctions and
invariant subspaces.  We first show that the T-SSD matrices capture
the maximal Koopman-invariant subspaces in the span of the original
dictionary~$D$.

\begin{theorem}\longthmtitle{T-SSD Matrices Capture the Maximal
    Koopman-Invariant
    Subspace}\label{t:tssd-contain-maximal-Koopman-invariant}
  Let $\Ic_{\max}$ denote the maximal Koopman-invariant subspace in
  $\Span(D)$ and let $\cmax$ be a full-column rank matrix such that
  $D(\cdot) \cmax$ spans $ \Ic_{\max}$ (if $\Ic_{\max}=\{0\}$, we set
  $\cmax = 0$).  Then, for any  $\epsilon \in [0,1]$,
  \begin{align*}
    \range(\cmax) \subseteq \range(C_i), \quad \forall i \in
    \{0,\ldots,L\},
  \end{align*}
  where $L$ and $C_i$ denote, respectively, the termination step and
  the $i$th internal matrix of T-SSD.
\end{theorem}
\begin{pf}
  The result holds trivially if $\Ic_{\max}=\{0\}$.  For the case
  $\Ic_{\max} \neq \{0\}$, we reason by induction. For $i = 0$,
  columns of $C_0$ span the whole space. Hence, $\range(\cmax)
  \subseteq \range(C_0)$. Next, assume $\range(\cmax) \subseteq
  \range(C_i)$ for $i \in \{0,1,\ldots,L-1\}$ and let us prove
  $\range(\cmax) \subseteq \range(C_{i+1})$.  The invariance of $
  \Ic_{\max}$ implies that $ \range(\dx \cmax) = \range(\dy \cmax)$.
  Using the definition of matrices $A_0, B_0$ in
  Algorithm~\ref{algo:tssd}, this can be equivalently written as
  $ \range(A_0 \cmax) = \range(B_0 \cmax)$.  Since $\range(\cmax)
  \subseteq \range(C_i)$, using Lemma~\ref{l:product-subspace}, we
  deduce
  \begin{align*}
    \range(A_0 \cmax) &\subseteq \range(A_0 C_i), \quad \range(B_0
    \cmax) \subseteq \range(B_0 C_i).
  \end{align*}
  Hence, $ \Pc_{A_0 C_i} w = w = \Pc_{B_0 C_i} w$, for all $ w \in
  \range(A_0 \cmax) = \range(B_0 \cmax)$, or equivalently,
  \begin{multline}\label{eq:proj-diff-zero}
    \| \Pc_{A_0 C_i} w - \Pc_{B_0 C_i} w \|_2 =0,
    \\
    \forall w \in \range(A_0 \cmax) = \range(B_0 \cmax).
  \end{multline}
  Now, noting that $A_i =A_0 C_i$ and $B_i = B_0 C_i$, one can use
  Step~\ref{algotssd:G} of Algorithm~\ref{algo:tssd} and write $
  G_{i+1} v = \Pc_{A_0 C_i} v - \Pc_{B_0 C_i} v$, for all $v \in
  \real^N$.  This, combined with~\eqref{eq:proj-diff-zero}, yields $
  \range(A_0 \cmax) = \range(B_0 \cmax) \subseteq
  \operatorname{null}(G_{i+1})$.  Therefore, since the eigenvectors of
  $G_{i+1}$ with zero eigenvalue span $\operatorname{null}(G_{i+1})$,
  we deduce from Step~\ref{algotssd:Vi},
  \begin{align}\label{eq:invariant-sub-in-v}
    \range(A_0 \cmax) = \range(B_0 \cmax) \subseteq \range(V_{i+1}).
  \end{align}
  Based on the induction hypothesis $\range(\cmax) \subset \range(C_i)$,
  and noting that $\cmax$ and $C_i$ have full column rank ($\cmax $ by
  definition and $C_i$ from Lemma~\ref{l:tssd-basic-properties}(b)),
  there exits a full-column rank matrix $F_i$ such that
  \begin{align}\label{eq:ci-cbar-ralationship}
    \cmax = C_i F_i .
  \end{align}
  Now,
  using~\eqref{eq:invariant-sub-in-v}-\eqref{eq:ci-cbar-ralationship},
  in conjunction with Proposition~\ref{p:symmetric-intersection}(c),
  we deduce $\range(F_i) \subseteq \range(E_{i+1})$. Consequently, one
  can use Lemma~\ref{l:product-subspace} and write $\range(\cmax) =
  \range(C_i F_i) \subseteq \range(C_i E_{i+1}) = \range(C_{i+1})$,
  concluding the proof. \qed
\end{pf}

Theorem~\ref{t:tssd-contain-maximal-Koopman-invariant} implies that
the subspace identified by T-SSD contains the maximal
Koopman-invariant subspace in~$\Span(D)$.

\begin{corollary}\longthmtitle{T-SSD Subspace Contains the Maximal
    Koopman-Invariant Subspace}\label{c:tssd-contains-max-invariant}
  Let $\Ic_{\max}$ be the maximal Koopman-invariant subspace in
  $\Span(D)$. Given $\epsilon \in [0,1]$, let $\ctssd$ and $\dtssd$
  be the output and dictionary identified by T-SSD according
  to~\eqref{eq:ctssd}-\eqref{eq:reduced-dictionary}. Then, $\Ic_{\max}
  \subseteq \Span(\dtssd)$.
\end{corollary}

The next result shows that the eigendecomposition of $\Ktssd$ captures
all Koopman eigenfunctions (and corresponding eigenvalues) in the span
of the original dictionary.

\begin{proposition}\longthmtitle{$\Ktssd$ Captures All Koopman
    Eigenfunctions in $\Span(D)$}\label{p:tssd-captures-eigenfunction}
  Let $\phi$ be a Koopman eigenfunction in $\Span(D)$ with eigenvalue
  $\lambda$.  For $\epsilon \in [0,1]$, let $\Ktssd$
  in~\eqref{eq:Ktssd} be the T-SSD predictor matrix. Then, $\phi \in
  \Span(\dtssd)$ and there exists $w$ with $\Ktssd w = \lambda w$
  such that $\phi(\cdot) = \dtssd(\cdot) w$.
\end{proposition}
\begin{pf}
  Note that $\phi$ must belong to the maximal Koopman-invariant
  subspace $\Ic_{\max}$ in $\Span(D)$ which, from
  Corollary~\ref{c:tssd-contains-max-invariant}, is included in $
  \Span(\dtssd) = \Span(D(\cdot) \ctssd) $. Therefore, there exists
  a complex vector $w$ of appropriate size such that $\phi(\cdot) =
  \dtssd(\cdot) w$.  Using now the interpretation of $\Ktssd$ as
  the EDMD solution with dictionary $\dtssd$ and data $X$, $Y$, it
  follows from~\citep[Lemma 4.1(b)]{MH-JC:22-tac} that $\Ktssd w =
  \lambda w$, as claimed. \qed
\end{pf}

Proposition~\ref{p:tssd-captures-eigenfunction} states that all
eigenfunctions in the span of the original dictionary $D$ belong to
the set of approximated eigenfunctions calculated with the dictionary $\dtssd$
defined by T-SSD.

\begin{remark}\longthmtitle{Monotonicity of T-SSD Subspaces} {\rm In
    general, the output of the T-SSD algorithm is not monotonic as a
    function of the design parameter $\epsilon$, i.e., it might be the
    case that $\Span(\dtssd^{\epsilon_1}) \not \subset
    \Span(\dtssd^{\epsilon_2})$ for $\epsilon_1 < \epsilon_2$.  In
    case monotonicity is desirable for a specific application, one can
    modify Step~\ref{algotssd:Vi} of Algorithm~\ref{algo:tssd} to
    remove only the eigenvector with the largest eigenvalue (in
    magnitude) that exceeds the desired accuracy level. This
    modification ensures monotonicity in $\epsilon$ at the cost of
    requiring the modified algorithm more iterations to terminate.
    All the results remain valid for the modified version of the
    algorithm.  \oprocend }
\end{remark}

\section{EDMD and SSD are Special Cases of T-SSD}
Consistent with our assertion that T-SSD balances accuracy and
expressiveness, here we show that EDMD on the original dictionary
(maximum expressiveness) corresponds to T-SSD with $\epsilon =1$ and
that SSD (maximum accuracy) corresponds to T-SSD with $\epsilon =
0${\footnote{We refer to $\Kssd$ and $\Ktssd$ as SSD and T-SSD Koopman
    approximations, which can be calculated by applying EDMD on
    dictionaries identified by SSD and T-SSD respectively.}.  We start
  by showing an important property of EDMD.

\begin{lemma}\longthmtitle{Linear Transformations Do not Change the
    Information Extracted by EDMD}\label{l:EDMD-coordinate-change}
  Let $D_1$ and $D_2$ be two dictionaries such that $D_1(\cdot) =
  D_2(\cdot) R$, with $R$ invertible.  Let
  Assumption~\ref{a:full-rank} hold for both dictionaries given data
  matrices $X$ and $Y$. Define
  \begin{align*}
    \Kedmd^1 & = \EDMD{D_1}{X}{Y} = D_1(X)^\dagger D_1(Y),
    \\
    \Kedmd^2 & = \EDMD{D_2}{X}{Y} =D_2(X)^\dagger D_2(Y).
  \end{align*}
  Then, $\Kedmd^1 = R^{-1} \Kedmd^2 R$. Therefore, $(\lambda,v)$ is an
  eigenpair of $\Kedmd^1$ if and only if $(\lambda,R v)$ is an
  eigenpair of $\Kedmd^2$.
\end{lemma}
\begin{pf}
  Based on Assumption~\ref{a:full-rank}, we have $\Kedmd^1 = \big(
  D_1(X)^T D_1(X) \big)^{-1} D_1(X)^T D_1(Y)$.  Using $D_1(\cdot) =
  D_2(\cdot) R$,
  \begin{align*}
    \Kedmd^1 &= \big(R^T D_2(X)^T D_2(X) R \big)^{-1} R^T D_2(X)^T
    D_2(Y) R
    \\
    &= R^{-1} \big( D_2(X)^T D_2(X) \big)^{-1} D_2(X)^T D_2(Y) R
    \\
    &= R^{-1} D_2(X)^\dagger D_2(Y) R = R^{-1} \Kedmd^2 R . 
  \end{align*}
  The rest follows from the properties of similarity transformations. \qed
\end{pf}

Lemma~\ref{l:EDMD-coordinate-change} states that the dynamical
information captured by the EDMD algorithm remains the same under
linear transformation of the dictionary. Note that the result does not
require the dictionaries to span a Koopman-invariant subspace.
We are ready to show that EDMD applied to the original dictionary
is a special case of T-SSD.

\begin{theorem}\longthmtitle{EDMD is a Special Case of T-SSD with
    $\epsilon =1$}\label{t:EDMD-spacialcase-TSSD}
  For $\epsilon =1$, let $\dtssd$ be the dictionary identified by
  T-SSD, cf.~\eqref{eq:reduced-dictionary}. Then, $\Span(\dtssd) =
  \Span(D)$, and $\Ktssd = \EDMD{\dtssd}{X}{Y} $ and $\Kedmd =
  \EDMD{D}{X}{Y} $ are similar and capture the same dynamical
  information.
\end{theorem}
\begin{pf}
  In the first iteration of Algorithm~\ref{algo:tssd}, one can
  use Step~\ref{algotssd:G} and the definition of $A_0$ and $B_0$ to
  write
  \begin{align*}
    G_1 = A_0 A_0^\dagger - B_0 B_0^\dagger = \Pc_{\dx} - \Pc_{\dy} .
  \end{align*}
  Since $G_1$ is symmetric, its eigenvalues are real. Moreover, they
  belong to $[-1,1]$, see e.g.~\citep[Lemma
  1]{WNA-EJH-GET:85}. Therefore, since $\epsilon = 1$, using
  Step~\ref{algotssd:Vi}, one can deduce that the columns of $V_1$
  span $\real^N$. As a result,
  \begin{align*}
    \range(\dx) = \range(A_0) = \range(A_0 I_{N_d}) \subseteq
    \range(V_1) = \real^N,
    \\
    \range(\dy) = \range(B_0) = \range(B_0 I_{N_d}) \subseteq
    \range(V_1) = \real^N.
  \end{align*}
  This, combined with the maximality of $E_1$ defined in
  Step~\ref{algotssd:symmetric-intersection},
  cf. Proposition~\ref{p:symmetric-intersection}(c), implies $
  \range(I_{N_d}) \subseteq \range(E_1)$.  Hence, $E_1$ is nonzero and
  has full column rank
  (cf. Proposition~\ref{p:symmetric-intersection}(a)). As a result,
  nothing that $\rown(E_1) = N_d$ , we deduce that $E_1$ is a
  nonsingular square matrix. Therefore, $ \range(C_1) = \range(C_0
  E_1) = \real^{N_d}$.  This and the fact that $E_1$ is square mean
  that the condition in Step~\ref{algotssd:size-check} is met and the
  algorithm executes
  Steps~\ref{algotssd:complete}-\ref{algotssd:break-complete}. Consequently,
  $\ctssd = C_1$ is a nonsingular square matrix and $
  \Span(\dtssd(\cdot)) = \Span(D(\cdot) \ctssd) = \Span(D(\cdot))$, so
  $\dtssd$ is a (potentially different) basis for the space spanned by
  $D$. The rest of the statement follows from
  Lemma~\ref{l:EDMD-coordinate-change}.  \qed
\end{pf}

The SSD algorithm is also a special case of T-SSD.

\begin{theorem}\longthmtitle{SSD is a Special Case of T-SSD with
    $\epsilon =0$}\label{t:SSD-spacialcase-TSSD}
  Let $\dssd(\cdot)$ be the dictionary identified by SSD,
  cf.~\eqref{eq:ssd-dictionary}, and, for $\epsilon =0$, let $\dtssd$
  be the dictionary identified by T-SSD,
  cf.~\eqref{eq:reduced-dictionary}.  Then, $\Span(\dtssd) =
  \Span(\dssd)$, and $\Ktssd = \EDMD{\dtssd}{X}{Y} $ and $\Kssd =
  \EDMD{\dssd}{X}{Y} $ are similar and capture the same dynamical
  information.
\end{theorem}
\begin{pf}
  Since $\epsilon = 0$, Theorem~\ref{t:tssd-output-eps-apart} implies
  that $\range(\dx \ctssd)$ and $\range(\dy \ctssd)$ are
  0-apart. Therefore, from Lemma~\ref{l:zero-apart-equal}, $
  \range(\dx \ctssd) = \range(\dx \ctssd)$.  This, together with
  Theorem~\ref{t:ssd-convergence}(c), implies
  \begin{align}\label{eq:ctssd-sub-cssd}
    \range(\ctssd) \subseteq \range(\cssd).
  \end{align}
  If $\cssd = 0$, then $\ctssd = 0$, and the proof is complete.
  Suppose instead that $\cssd \neq 0$, with full column rank,
  cf.~Theorem~\ref{t:ssd-convergence}(a). We use induction to prove
  that $\range(\cssd) \subseteq \range(C_i)$, where $C_i$ is the
  internal matrix of the T-SSD algorithm for $i \in \zuntil{L}$ and
  $L$ is the iteration at which it terminates.  When $i = 0$, the
  columns of $C_0 =I_{N_d}$ span $\real^{N_d}$ and, therefore,
  $\range(\cssd) \subseteq \range(C_0)$.  Assume then that
  $\range(\cssd) \subseteq \range(C_i)$ for $i \in \{0,\ldots,L-1\}$,
  and let us prove that $\range(\cssd) \subseteq \range(C_{i+1})$.
	
  Based on Theorem~\ref{t:ssd-convergence}(b), we have $ \range(\dx
  \cssd) = \range(\dy \cssd)$.  This, together the definition of
  matrices $A_0, B_0$ in Algorithm~\ref{algo:tssd} and the fact
  that $\range(\cssd) \subseteq \range(C_i)$, yields
  \begin{align}\label{eq:proj-diff-zero-ssd}
    \Pc_{A_0 C_i} w = \Pc_{B_0 C_i} w = w ,
  \end{align}
  for all $w \in \range(A_0 \cssd) = \range(B_0 \cssd)$.  Now, since
  $A_i =A_0 C_i$ and $B_i = B_0 C_i$ at iteration $i+1$ of the T-SSD
  algorithm, $ G_{i+1} v = \Pc_{A_0 C_i} v - \Pc_{B_0 C_i} v$, for all
  $ v \in \real^N$.  This, together
  with~\eqref{eq:proj-diff-zero-ssd}, implies that $ \range(A_0 \cssd)
  = \range(B_0 \cssd) \subseteq \operatorname{null}(G_{i+1})$.  Since
  $\epsilon = 0$, from Step~\ref{algotssd:Vi} we know that $V_{i+1}$
  is a basis for $ \operatorname{null}(G_{i+1})$, and therefore
  \begin{align}\label{eq:invariant-sub-in-v-ssd}
    \range(A_0 \cssd) = \range(B_0 \cssd) \subseteq \range(V_{i+1}).
  \end{align}
  By the induction hypothesis $\range(\cssd) \subseteq
  \range(C_i)$. This, together with the fact that $\cssd$ and $C_i$
  have full column rank (the latter because of
  Lemma~\ref{l:tssd-basic-properties}(b)), implies that there exists a
  matrix $F_i$ with full column rank such that
  \begin{align}\label{eq:ci-cbar-ralationship-ssd}
    \cssd = C_i F_i .
  \end{align}
  Using
  now~\eqref{eq:invariant-sub-in-v-ssd}-\eqref{eq:ci-cbar-ralationship-ssd}
  together with the fact that $A_i =A_0 C_i$, $B_i =B_0 C_i $, one can
  invoke Proposition~\ref{p:symmetric-intersection}(c) to deduce that
  $\range(F_i) \subseteq \range(E_{i+1})$. Consequently,
  \begin{align*}
    \range(\cssd) = \range(C_i F_i) \subseteq \range(C_i E_{i+1}) = \range(C_{i+1}).
  \end{align*}
  Hence, the induction is complete and
  \begin{align}\label{eq:cssd-subset-ci}
    \range(\cssd) \subseteq \range(C_i), \, \forall i \in \until{L}.
  \end{align}
  Since $\cssd$ is nonzero and has full column rank, one can deduce
  that $C_L$ is nonzero and has full column rank as a result of
  Lemma~\ref{l:tssd-basic-properties}(c). Consequently, the T-SSD
  algorithm terminates by executing
  Steps~\ref{algotssd:complete}-\ref{algotssd:break-complete}. Therefore,
  $\ctssd = C_L$ and using~\eqref{eq:ctssd-sub-cssd}
  and~\eqref{eq:cssd-subset-ci}, we have $\range(\ctssd) =
  \range(\cssd)$ and consequently $\Span(\dtssd) = \Span(\dssd)$.
  The rest of the statement follows from
  Lemma~\ref{l:EDMD-coordinate-change}.  \qed
\end{pf}

It is worth mentioning that, when implementing T-SSD for $\epsilon
=0$, we have found it useful to set $\epsilon$ to be a small positive
number (instead of zero) to avoid complications by round-off errors.

\begin{remark}\longthmtitle{Dynamical Properties of T-SSD Subspace
    with $\epsilon =0$}
  {\rm Given Theorem~\ref{t:SSD-spacialcase-TSSD}, the subspace
    identified by T-SSD for $\epsilon=0$ enjoys important dynamical
    properties, cf. Section~\ref{sec:SSD}: under reasonable conditions
    on the density of data sampling,
    cf.\citep[Theorems~5.7-5.8]{MH-JC:22-tac}, the identified subspace
    is the maximal Koopman-invariant subspace in the span of the
    dictionary almost surely. Moreover, the eigenfunctions and
    predictors identified by T-SSD are almost surely exact.}
  \oprocend
\end{remark}

\section{Efficient Implementation of T-SSD}\label{sec:efficient_TSSD}

Here, we propose a modification to the implementation of the T-SSD
algorithm on digital computers to increase efficiency. This is based
on the following observation: a close look at the form of the matrix
$G_i \in \real^{N \times N}$ in Step~\ref{algotssd:G} of
  Algorithm~\ref{algo:tssd} as a difference of projections reveals
that its eigenvectors are either in or orthogonal to the subspace
$\range(A_{i-1})+ \range(B_{i-1})$, see e.g.,~\citep{WNA-EJH-GET:85}.
However, in Step~\ref{algotssd:symmetric-intersection}, the matrix
$E_i$ satisfies $\range(A_{i-1} E_i), \range(B_{i-1} E_i) \subseteq
\range(V_i)$. Hence, the $\symmintersection$ function filters out all
eigenvectors of $G_i$ that are orthogonal to $\range(A_{i-1}) +
\range(B_{i-1})$, i.e., these eigenvectors are never used. This is
despite the fact that, since generally $N \gg N_d$, such eigenvectors
form a majority of eigenvectors of $G_i$ (at least $N-2N_d$ out of
$N$).

This motivates us to seek a method that bypasses the calculation of
the unused eigenvectors of~$G_i$.  To achieve this goal, let $H_i$ be
a matrix such that
\begin{align}\label{eq:orthogonal-H}
  \range(H_i) := \range([A_{i-1},B_{i-1}]), \quad H_i^T H_i =
  I_{\coln(H_i)}.
\end{align}
The columns of $H_i$ form an orthonormal basis of $\range(A_{i-1}) +
\range(B_{i-1})$. One can calculate $H_i$ by applying the Gram–Schmidt
process, or other closely related orthogonal factorization method such
as QR decomposition (see e.g.~\citep{LNT-DB:97}), on
$[A_{i-1},B_{i-1}]$. The next result shows that the eigendecomposition
of the matrix $H_i^T G_i H_i$ completely captures the
eigendecomposition of $G_i$ on $\range(A_{i-1}) + \range(B_{i-1})$.

\begin{proposition}\longthmtitle{Eigenvectors of $H_i^T G_i H_i$
    Characterize All Eigenvectors of $G_i$ in $\range(A_{i-1}) +
    \range(B_{i-1})$}\label{p:H-eigendecomposition}
  Let $G_i$ as defined in Step~\ref{algotssd:G} of 
    Algorithm~\ref{algo:tssd}, and let
  $H_i$ satisfy~\eqref{eq:orthogonal-H}. Then, $w \in
  \cplx^{N_d}\setminus \{0\}$ is an eigenvector of $H_i^T G_i H_i$
  with eigenvalue $\lambda$ if and only if $v = H_i w$ is an
  eigenvector of $G_i$ with eigenvalue $\lambda$.
\end{proposition}
\begin{pf}
  $(\Leftarrow)$ By hypothesis, $G_i H_i w = \lambda H_i w$. Hence, $
  H_i^T G_i H_i w = \lambda H_i^T H_i w = \lambda w$ (where we have
  used~\eqref{eq:orthogonal-H}).
  
  $(\Rightarrow)$ By hypothesis, $H_i^T G_i H_i w = \lambda
  w$. Using~\eqref{eq:orthogonal-H}, this can be rewritten as
  \begin{align}\label{eq:H_i-null}
    H_i^T (G_i H_i w - \lambda H_i w)= 0.
  \end{align}
  By definition of $G_i$, we can write $ G_i H_i w = \Pc_{A_{i-1}}
  (H_i w) - \Pc_{B_{i-1}} (H_i w)$.  From~\eqref{eq:orthogonal-H}, we
  have $\range(A_{i-1}), \range(B_{i-1}) \subseteq \range(H_i)$.
  Since $\Pc_{A_{i-1}} (H_i w) \in \range(A_{i-1})$ and $
  \Pc_{B_{i-1}} (H_i w) \in \range(B_{i-1})$, we deduce that $G_i H_i
  w \in \range(H_i)$, and consequently, $G_i H_i w - \lambda H_i w \in
  \range(H_i)$. However, from~\eqref{eq:H_i-null}, $(G_i H_i w -
  \lambda H_i w) \in \operatorname{null}(H_i^T)$. Therefore, since
  $\range(H_i) \perp \operatorname{null}(H_i^T)$, we conclude $G_i H_i
  w - \lambda H_i w = 0$, as claimed.  \qed
\end{pf}

Based on Proposition~\ref{p:H-eigendecomposition}, we modify T-SSD to
achieve higher computational efficiency. Formally, the
\textbf{Efficient T-SSD} algorithm replaces Steps~\ref{algotssd:G}
and~\ref{algotssd:symmetric-intersection} of Algorithm~\ref{algo:tssd}
by
\begin{align*}
\text{\ref{algotssd:G}.a:} \quad & H_i \gets \basis([A_i,B_i])
  \\
  \text{\ref{algotssd:G}.b:} \quad & G_i \gets H_i^T (A_{i-1}
  A_{i-1}^\dagger - B_{i-1} B_{i-1}^\dagger) H_i
  \\
  \text{\ref{algotssd:symmetric-intersection}:} \quad & E_i \gets
  \symmintersection(H_i V_i, A_{i-1}, B_{i-1})
\end{align*}
These steps bypass the computation of the (unused) eigenvectors of
$G_i$ that are orthogonal to $\range(A_{i-1}) + \range(B_{i-1})$ in
the original T-SSD implementation.

\begin{remark}\longthmtitle{Computational Complexity of Efficient
    T-SSD}\label{r:efficient-tssd-complexity}
  {\rm Given $N$ data snapshots and $N_d$ dictionary functions, and
    considering the complexity of scalar operations as $O(1)$, the
    most time-consuming steps of Efficient T-SSD are calculating $H_i$
    in~\eqref{eq:orthogonal-H} and the null space calculations in the
    function $\symmintersection$, which can be done in $O(NN_d^2)$
    operations. Since the algorithm terminates after at most $N_d$
    iterations, cf. Proposition~\ref{p:finite-time-termination}, the
    overall complexity is $O(NN_d^3)$.  Compared to T-SSD,
    cf. Remark~\ref{r:tssd-complexity}, the efficient T-SSD algorithm
    provides a reduction of $O(N^2 N_d^{-2})$, leading to a drastic
    reduction in run time for typical situations, where $N \gg N_d$.
    \oprocend }
\end{remark}

\section{Simulation Results}\label{sec:simulations}
Here, we illustrate the effectiveness of our proposed methods using
three examples.

\subsection{Hopf Normal Form}
Consider the system~\citep{SLB-JLP-JNK:16,JEM-MM:12} on $\Mc=
[-2,2]^2$,
\begin{align}\label{eq:Hopf}
&\dot{x_1} = x_1 + 2 x_2 - x_1 (x_1^2 + x_2^2),
\nonumber \\
&\dot{x_2} = -2 x_1 + x_2 - x_2 (x_1^2 + x_2^2),
\end{align}
with state $x=[x_1, x_2]^T$, which admits an attractive periodic
orbit. We consider the discretized version of~\eqref{eq:Hopf} with
time step $\Delta t= 0.01s$ and gather $N = 10^4$ data snapshots in
matrices $X$ and $Y$, with initial conditions uniformly selected
from~$\Mc$. We consider the space of all polynomials up to degree
  $10$ spanned by all the $N_d = 66$ distinct monomials in the form of
  $\prod_{i=1}^{10} \alpha_i$, with $\alpha_i \in \{1, x_1, x_2\}$ for
  $i \in \until{10}$. To ensure resilience against machine precision
  errors and providing informative representations, we choose a
  dictionary $D$ with $N_d = 66$ functions such that the columns of
  $\dx$ are orthonormal\footnote{This dictionary can be found by
      first forming a dictionary comprised of the monomials and then
      performing a linear transformation on the dictionary to make the
      columns orthonormal. The linear transformation does not impact
      the captured dynamical information
      (cf. Lemma~\ref{l:EDMD-coordinate-change}).}}.

We implement the Efficient T-SSD algorithm,
cf.~Section~\ref{sec:efficient_TSSD}, with $\epsilon \in \{0.02, 0.05,
0.1, 0.15, 0.2\}$. Table~\ref{table:dimension-vs-epsilon-Hopf} shows
the dimension of the identified dictionary, $\dtssd$, versus the
value of the design parameter $\epsilon$. For $\epsilon = 0.2$, T-SSD
identifies the original dictionary, certifying that the range spaces
of $\dx$ and $\dy$ are $0.2$-apart. On the other hand, the
one-dimensional subspace identified by $\epsilon = 0.02$ is in fact
the maximal Koopman-invariant subspace of $\Span(D)$, spanned by the
trivial eigenfunction $\phi(x) \equiv 1$ with eigenvalue $\lambda =1$.

{
  \renewcommand{\arraystretch}{1.5}
  \begin{table}[htb]
    \centering
    \caption{Dimension of subspace identified by Efficient T-SSD vs $\epsilon$ for~\eqref{eq:Hopf}.} \label{table:dimension-vs-epsilon-Hopf}
    \begin{tabular}[htb]{ | c || c | c | c | c | c | }
      \hline
      \textbf{$\boldsymbol{\epsilon}$}	               				   & 0.02  & 0.05  & 0.10  & 0.15  & 0.20  \\ \hline %
      \textbf{$\mathbf{\boldsymbol{\dim \dtssd}}$}           & 1        & 6        &    8  & 16     & 66      \\ \hline %
    \end{tabular}
  \end{table}
}

\begin{figure}[htb]
  \centering
  {\includegraphics[width=.45\linewidth]{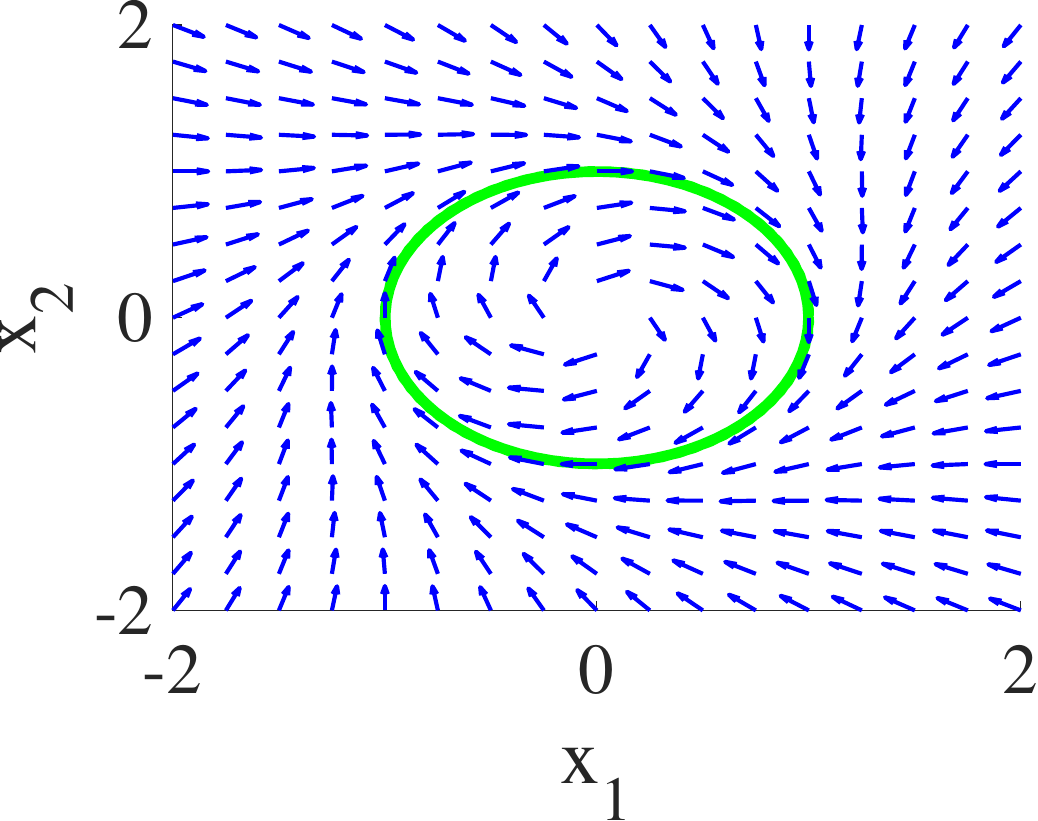}}
  {\includegraphics[width=.45\linewidth]{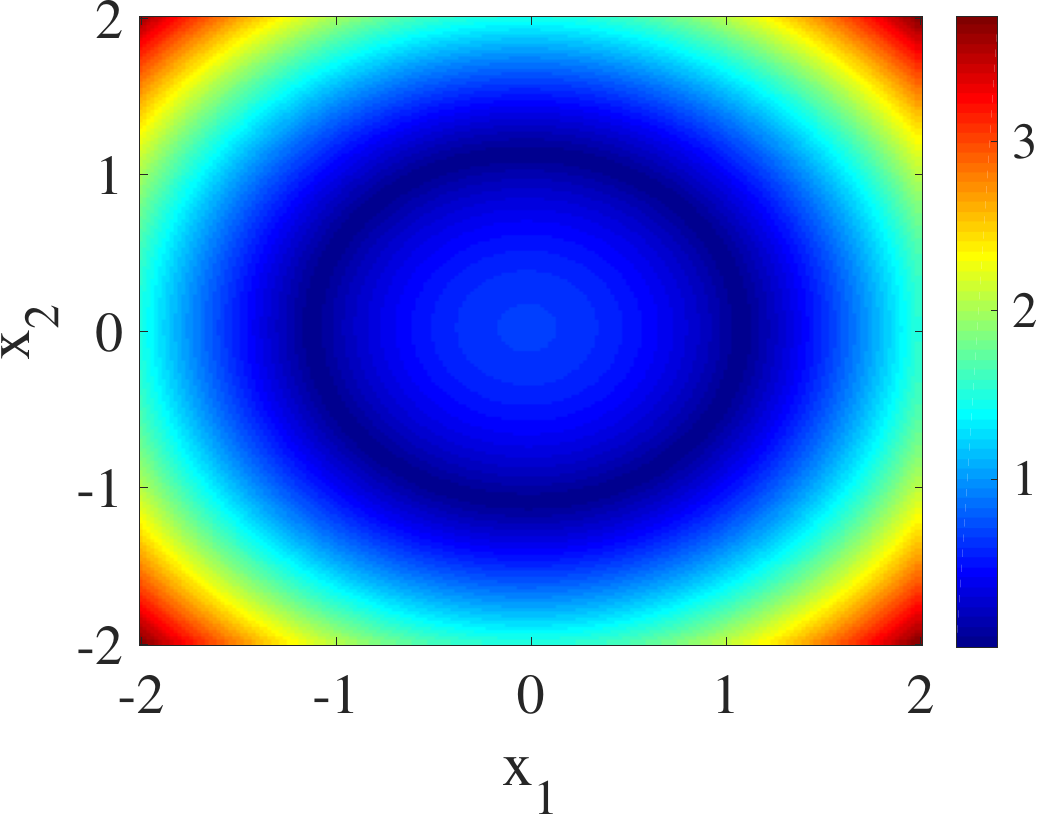}}
  \caption{Vector field and periodic orbit of system~\eqref{eq:Hopf}
    (left) and the absolute value of eigenfunction with eigenvalue
    $\lambda = 0.9066$ (right).}\label{fig:real_eigenfunction_Hopf}
  \vspace*{-1ex}
\end{figure}

To demonstrate the effectiveness of the T-SSD algorithm in
approximating Koopman eigenfunctions and invariant subspaces, we focus
on the subspace identified with $\epsilon = 0.05$.  In accordance with
Proposition~\ref{p:tssd-captures-eigenfunction}, T-SSD identifies the
trivial eigenfunction $\phi(x) \equiv 1$ spanning the maximal
Koopman-invariant subspace of $\Span(D)$.  T-SSD also approximates
another real-valued eigenfunction with eigenvalue $\lambda = 0.9066$,
whose absolute value is illustrated in
Figure~\ref{fig:real_eigenfunction_Hopf}(right). Given that $|0.9066|
< 1$, this eigenfunction predicts the existence of a forward invariant
set (the periodic orbit in
Figure~\ref{fig:real_eigenfunction_Hopf}(left)) at its zero-level set.

\begin{figure}[htb]
  \centering
  {\includegraphics[width=.45\linewidth]{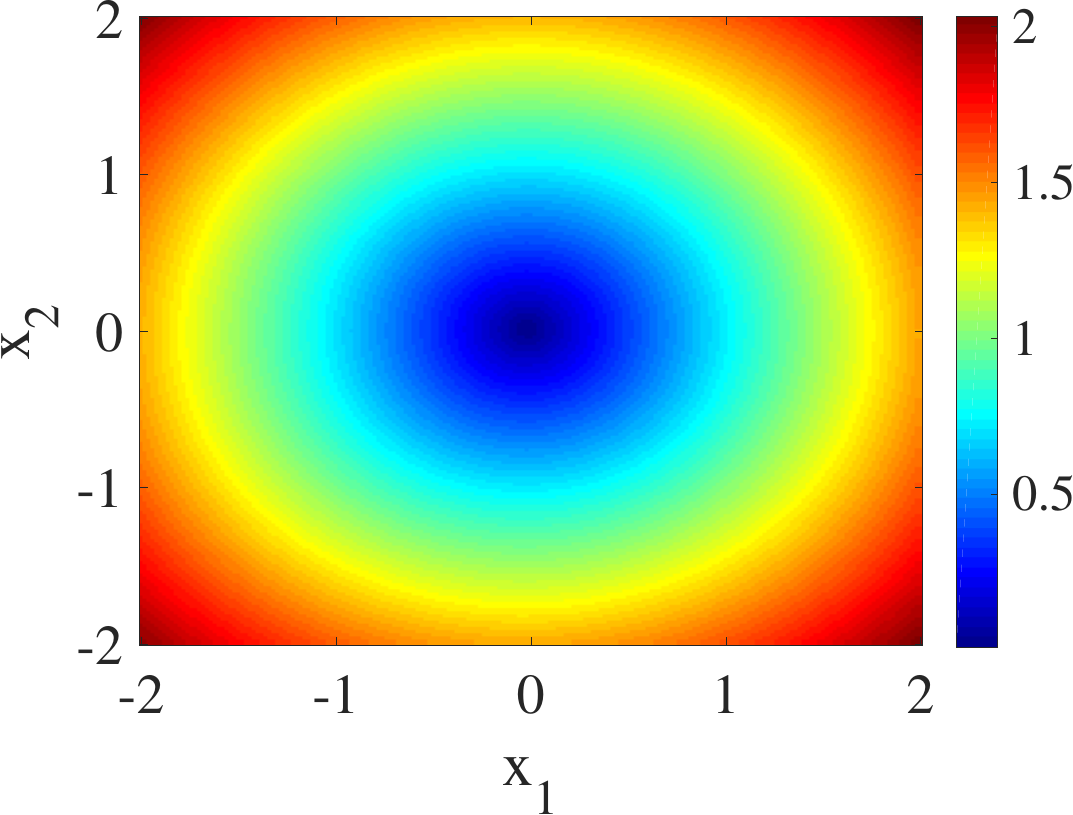}}
  {\includegraphics[width=.45\linewidth]{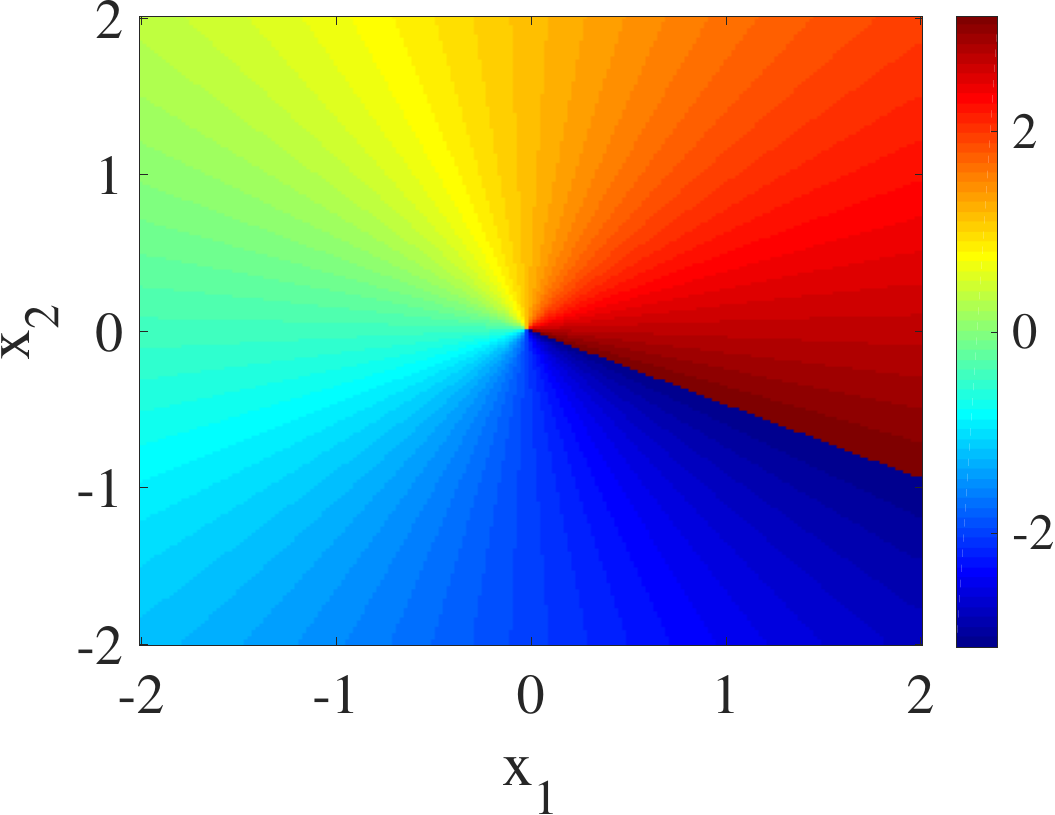}}
  \caption{Absolute value (left) and phase (right) of the
    eigenfunction with eigenvalue $\lambda = 0.9938 + 0.0195j$ for~\eqref{eq:Hopf}.}\label{fig:oscillatory-eigenfunction_Hopf}
\vspace*{-1ex}
\end{figure}

In addition, T-SSD also identifies two pairs of complex
eigenfunctions. For space reasons, we only show in
Figure~\ref{fig:oscillatory-eigenfunction_Hopf} one eigenfunction with
eigenvalue $\lambda = 0.9938 + 0.0195j$ (the one closest to the unit
circle).  Its phase characterizes the oscillation of the trajectories
in the state space.

\begin{figure}[htb]
  \centering 
  {\includegraphics[width=.48\linewidth]{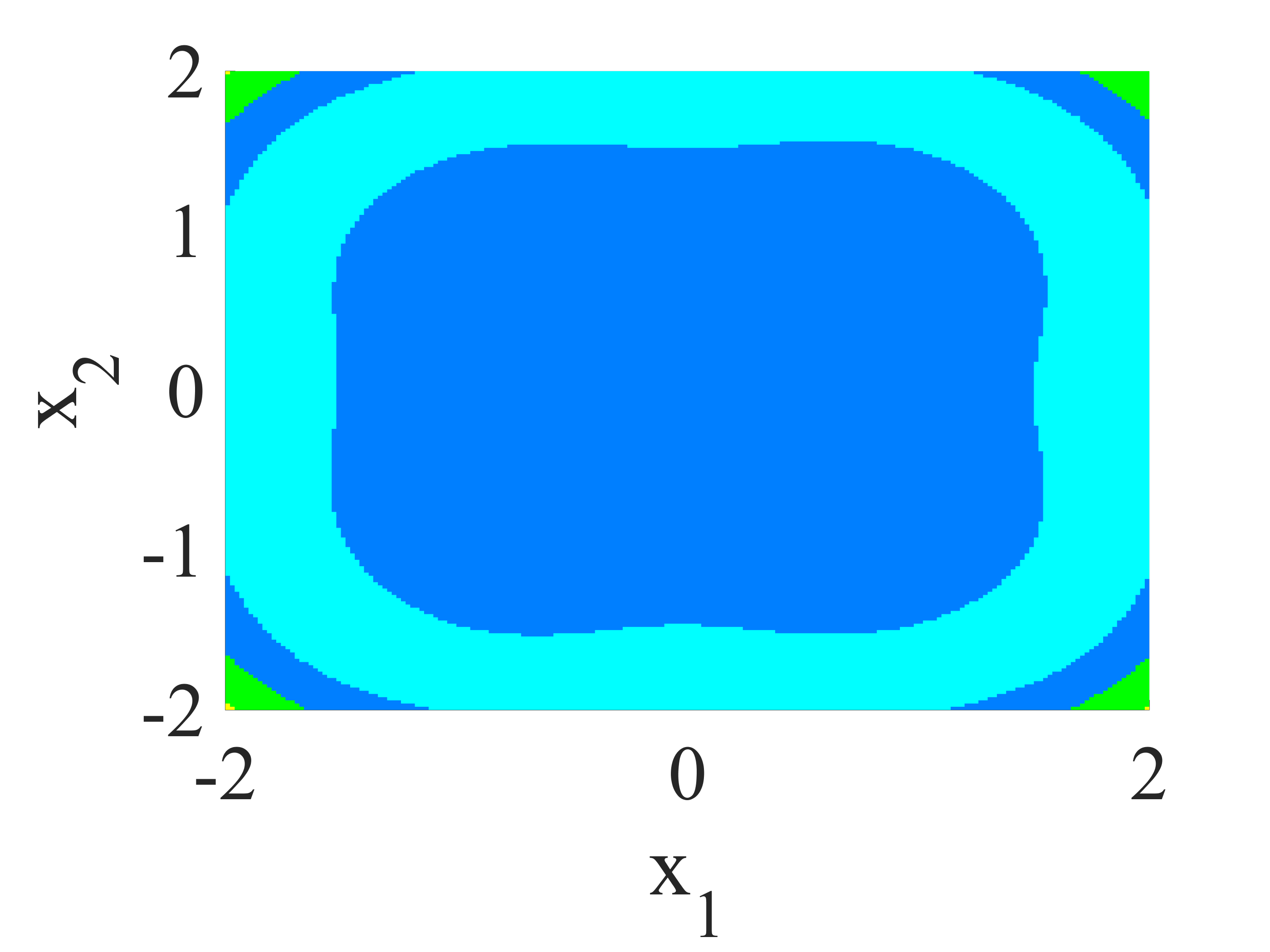}}
  {\includegraphics[width=.48\linewidth]{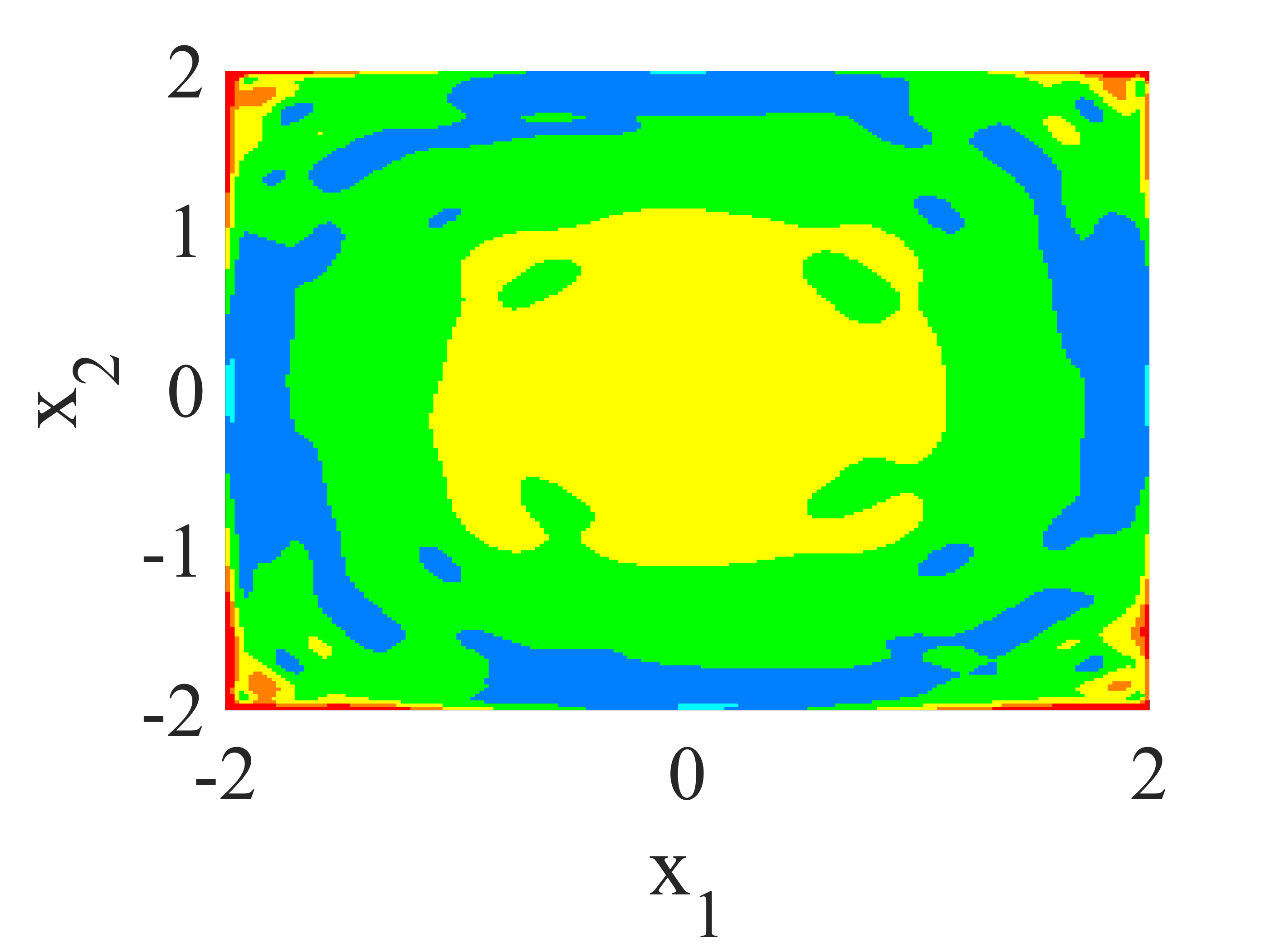}}
  \\
  {\includegraphics[width=.98\linewidth]{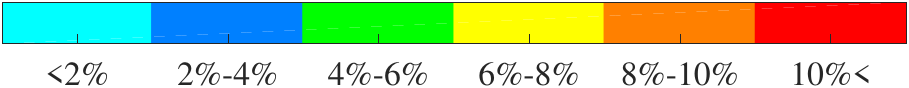}}
  \caption{Relative linear prediction error for dictionary identified
    by T-SSD ($\epsilon = 0.05$) 
    (left) and the original dictionary (right for~\eqref{eq:Hopf}.}\label{fig:relative_error_Hopf} 
  \vspace*{-1ex}
\end{figure}

To illustrate the efficacy of our algorithm regarding the prediction
accuracy of the dictionary, we consider the relative linear prediction
error associated with a dictionary $\Dc$ at point $x$ given data
snapshot matrices $X$ and $Y$ defined by
\begin{align}\label{eq:relative-linear-prediction-error}
  E_{\operatorname{relative}}(x) := \frac{\| \Dc \circ T(x) - \Dc(x)
    K\|_2}{\|\Dc \circ T (x)\|_2 } \times 100,
\end{align}
where $K = \EDMD{\Dc}{X}{Y}$. Figure~\ref{fig:relative_error_Hopf}
compares this error on the state space~$\Mc$ for the dictionary
$\dtssd$ identified by T-SSD with $\epsilon =0.05$ and for the
original dictionary~$D$. This error is evaluated at points other than
the training data~$X$ and clearly shows the advantage of $\dtssd$ over
the original dictionary both in prediction errors and capturing the
radial symmetry of the vector field.

Noting that the error in~\eqref{eq:relative-linear-prediction-error}
depends on the dictionary and does not provide information about the
subspace it spans (or the individual members of the subspace), we also
consider the latter. For this reason, we use the data sampling
strategy used earlier to build a test data set denoted by snapshot
matrices $\Xtest$ and $\Ytest$ with $N_{\operatorname{test}} = 10^4$
samples. Given a dictionary $\Dc$, we evaluate the invariance
proximity of $\Span(\Dc)$ as the smallest $\epsilon$ such that
$\range(\Dc(\Xtest))$ and $\range(\Dc(\Ytest))$ are $\epsilon$-apart.
This data-driven measure is equivalent to the maximum relative root
mean square error for a function in the span of a given dictionary
$\Dc$ on the test data defined as
  \begin{align}\label{eq:rrmse_max}
    &\operatorname{RRMSE_{\max}}(\Dc,\Xtest,\Ytest)
    \nonumber \\
    &= \max_{f \in \Span(\Dc)} \frac{\sqrt {\frac{1}{N_t}
        \sum_{i=1}^{N_t} | \Kc f (x_i) - \Pf_{\Kc f}^{\tssd}|^2 } }{
      \sqrt {\frac{1}{N_t} \sum_{i=1}^{N_t} | \Kc f (x_i)|^2 } },
  \end{align}
  where $x_i^T$ and $y_i^T$ correspond to the $i$th rows of $\Xtest$
  and $\Ytest$ respectively, $y_i = T(x_i)$, and $T$ is the map
  defining the dynamics~\eqref{eq:Hopf}. The predictor $\Pf_{\Kc f}^{\tssd}$ is
  defined in~\eqref{eq:predictor-tssd} and calculated based on
  the test data. It is important to note that the evaluation of the
  error in~\eqref{eq:rrmse_max} goes beyond the assumptions of
  Theorem~\ref{t:rrms-bound}, since the dictionary is identified with
  the original data $X,Y$, but the error is evaluated on the test data
  $\Xtest, \Ytest$ (instead, the guarantee of
  Theorem~\ref{t:rrms-bound} are only valid when the error is
  evaluated on the original data). Following the reasoning in the proof
  of Theorem~\ref{t:rrms-bound} and Definition~\ref{def:eps-apart},
  one can analytically show that
  \begin{align*}
    \operatorname{RRMSE_{\max}}(\Dc,\Xtest,\Ytest) = \lambda_{\max}
    (\Pc_{\Dc(\Xtest)} - \Pc_{\Dc(\Ytest)}),
  \end{align*}
  where $\lambda_{\max}$ denotes the largest eigenvalue of the
  argument. Table~\ref{table:RRMSEmax-vs-epsilon-Hopf} shows the
  maximum relative root mean square error for the subspaces identified
  by T-SSD given different values of~$\epsilon$. According to
  Table~\ref{table:RRMSEmax-vs-epsilon-Hopf}, despite the
    fact that we have used different data for identification and
    evaluation, the error on the test data satisfies the upper bound
  accuracy requirement enforced by the accuracy parameter~$\epsilon$ .

  {	
	\renewcommand{\arraystretch}{1.5}
	\begin{table}[htb]
		\centering
		\caption{Maximum Relative Root Mean Square Error vs $\epsilon$ for~\eqref{eq:Hopf}.} \label{table:RRMSEmax-vs-epsilon-Hopf}
		
		\begin{tabular}[htb]{ | c || c | c | c | c | c | }
			\hline
			\textbf{$\boldsymbol{\epsilon}$}	               				   & 0.02  & 0.05  & 0.10  & 0.15  & 0.20  \\ \hline %
			\textbf{$\operatorname{RRMSE_{\max}}$}           & $\sim 0$        & 0.037       & 0.100 & 0.115    &   0.185      \\ \hline %
		\end{tabular}
	
	\end{table}
}

\subsection{Duffing System}
Consider the Duffing system~\citep{MOW-IGK-CWR:15} on $\Mc= [-2,2]^2$,
\begin{align}\label{eq:Duffing}
  \dot{x_1} & = x_2 , \nonumber
  \\
  \dot{x_2} & = -0.5x_2 +x_1 (1-x_1^2),
\end{align}
with state $x=[x_1, x_2]^T$, which has an unstable equilibrium at the
origin and two asymptotically stable equilibria at $(-1,0)$ and
$(1,0)$. We consider the discretized version of~\eqref{eq:Duffing}
with time step $\Delta t= 0.02s$ and gather $N = 10^4$ data snapshots
in matrices $X$ and $Y$ from $5000$ trajectories with length equal to
two time steps and initial conditions uniformly selected
from~$\Mc$. Similarly to the previous example, we use a
  dictionary $D$ with $N_d = 66$ elements spanning the space of all
  polynomials up to degree $10$ such that the columns of $\dx$ are
  orthonormal.

We apply the Efficient T-SSD algorithm,
cf.~Section~\ref{sec:efficient_TSSD}, with $\epsilon \in \{0.01, 0.02,
0.08, 0.14, 0.2,
0.26\}$. Table~\ref{table:dimension-vs-epsilon-Duffing} shows the
dimension of the identified dictionary, $\dtssd$, versus the value of
the design parameter~$\epsilon$. For $\epsilon = 0.26$, T-SSD
identifies the original dictionary, certifying that the range spaces
of $\dx$ and $\dy$ are $0.26$-apart. On the other hand, the
one-dimensional subspace identified by $\epsilon = 0.01$ is in fact
the maximal Koopman-invariant subspace of $\Span(D)$, spanned by the
trivial eigenfunction $\phi(x) \equiv 1$ with eigenvalue $\lambda =1$.

{
\renewcommand{\arraystretch}{1.5}
\begin{table}[htb]
  \centering
  \caption{Dimension of subspace identified by Efficient T-SSD vs
    $\epsilon$ for~\eqref{eq:Duffing}.} \label{table:dimension-vs-epsilon-Duffing} 
  \begin{tabular}[htb]{ | c || c | c | c | c | c | c | }
    \hline
    \textbf{$\boldsymbol{\epsilon}$}	               				   & 0.01  & 0.02  & 0.08  & 0.14  & 0.20 & 0.26  \\ \hline %
    \textbf{$\mathbf{\boldsymbol{\dim \dtssd}}$}           & 1        & 2        &  20    & 44     & 58   &66     \\ \hline %
  \end{tabular}
\end{table}
}

\begin{figure}[htb]
  \centering 
  {\includegraphics[width=.45\linewidth]{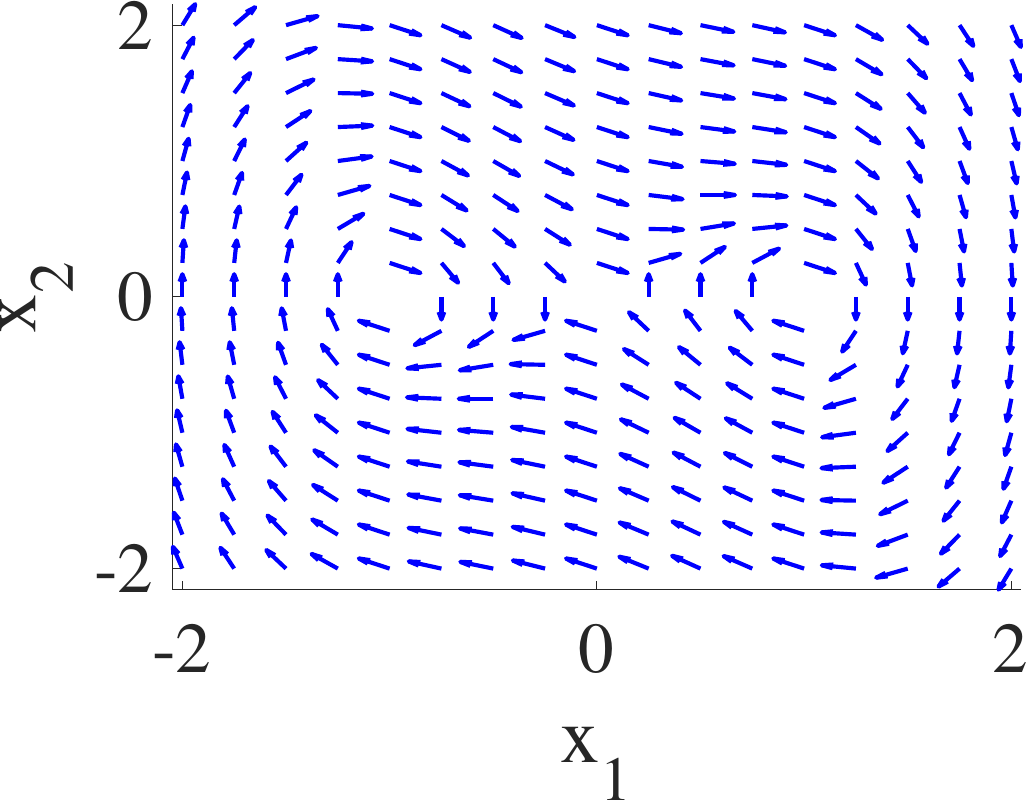}}
  {\includegraphics[width=.45\linewidth]{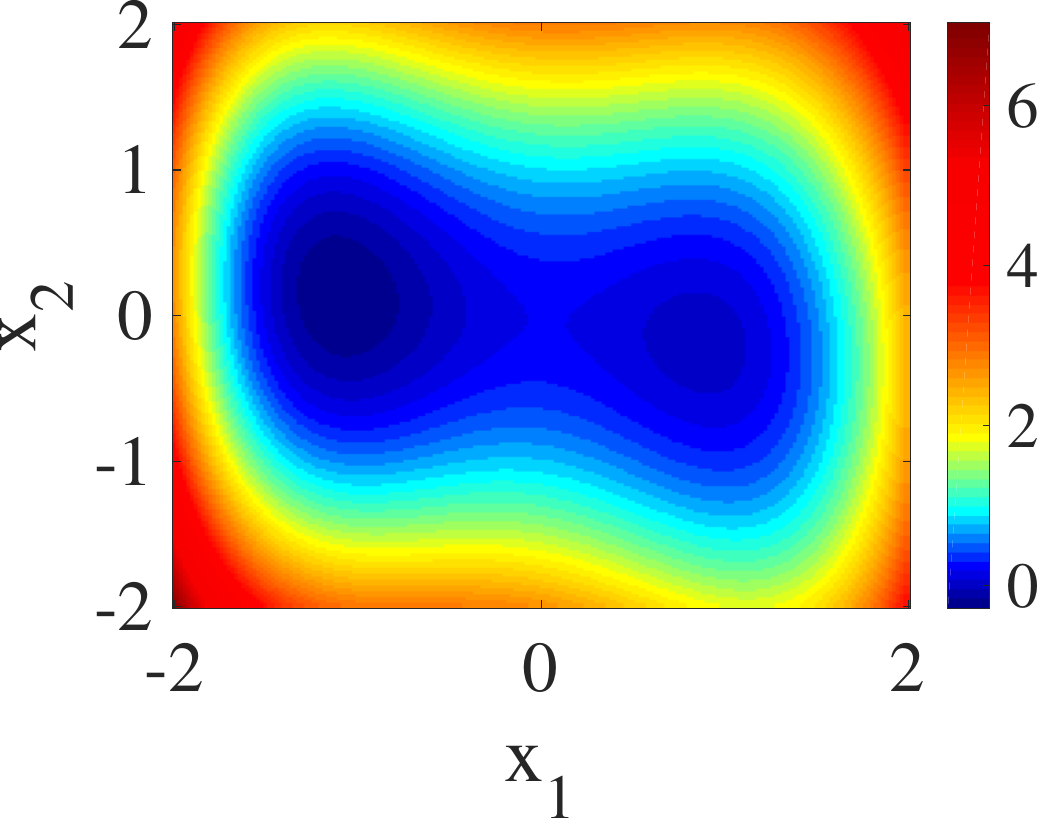}}
  \caption{Vector field (left) and eigenfunction with eigenvalue
    $\lambda = 0.9839$ (right) for~\eqref{eq:Duffing}.}\label{fig:real_eigenfunction_Duffing}
  \vspace*{-1ex}
\end{figure}

To demonstrate the effectiveness of the T-SSD algorithm in
approximating Koopman eigenfunctions and invariant subspaces, we focus
on the subspace identified with $\epsilon = 0.02$. Consistent with
Proposition~\ref{p:tssd-captures-eigenfunction}, T-SSD identifies the
trivial eigenfunction $\phi(x) \equiv 1$ spanning the maximal
Koopman-invariant subspace of $\Span(D)$.  T-SSD also approximates
another real-valued eigenfunction with eigenvalue $\lambda = 0.9839$
depicted in Figure~\ref{fig:real_eigenfunction_Duffing}(right), which
clearly captures the attractiveness of the asymptotically stable
equilibria and the general behavior of the vector field depicted in
Figure~\ref{fig:real_eigenfunction_Duffing}(left).

\begin{figure}[htb]
  \centering 
  {\includegraphics[width=.48\linewidth]{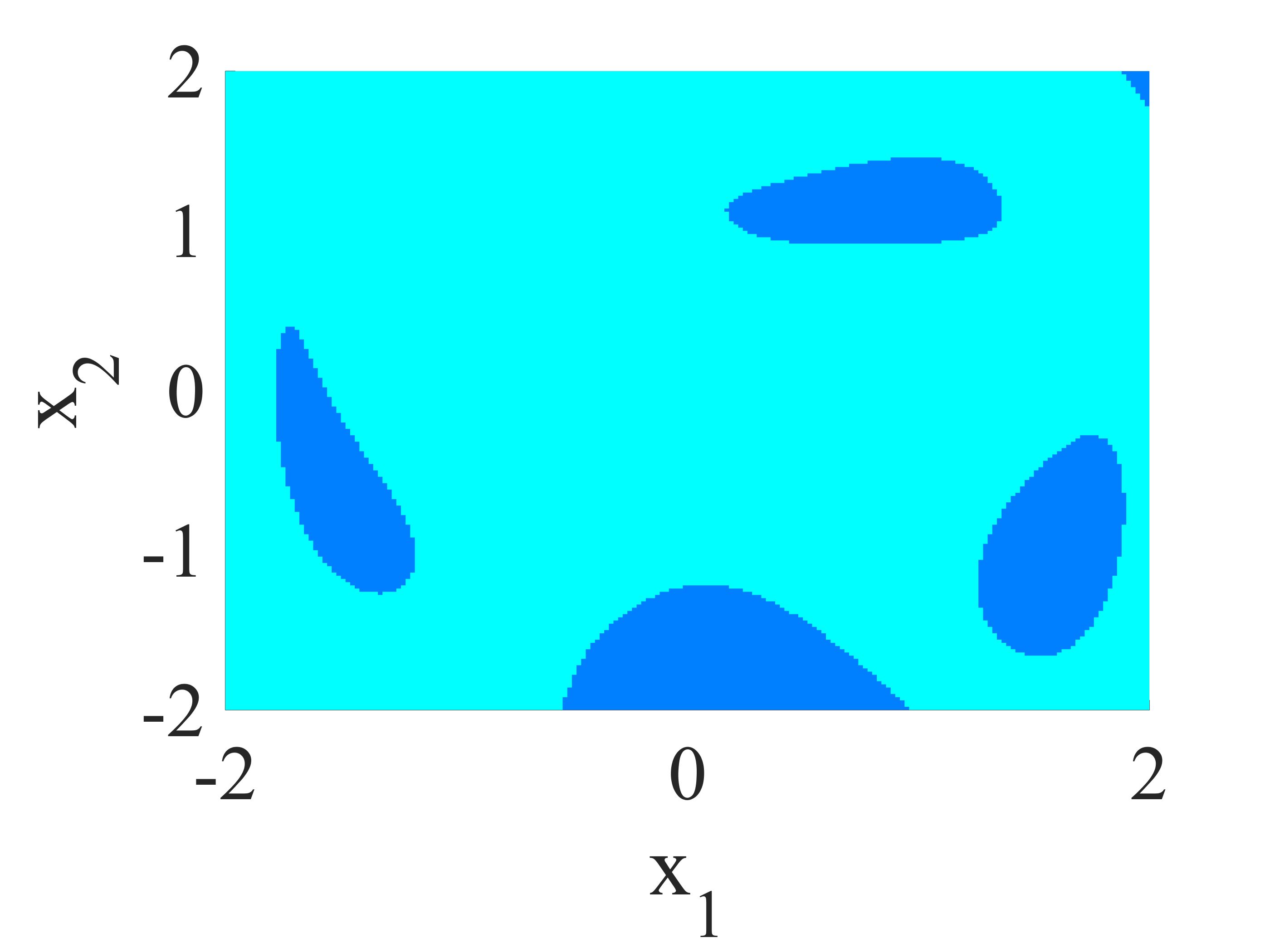}}
  {\includegraphics[width=.48\linewidth]{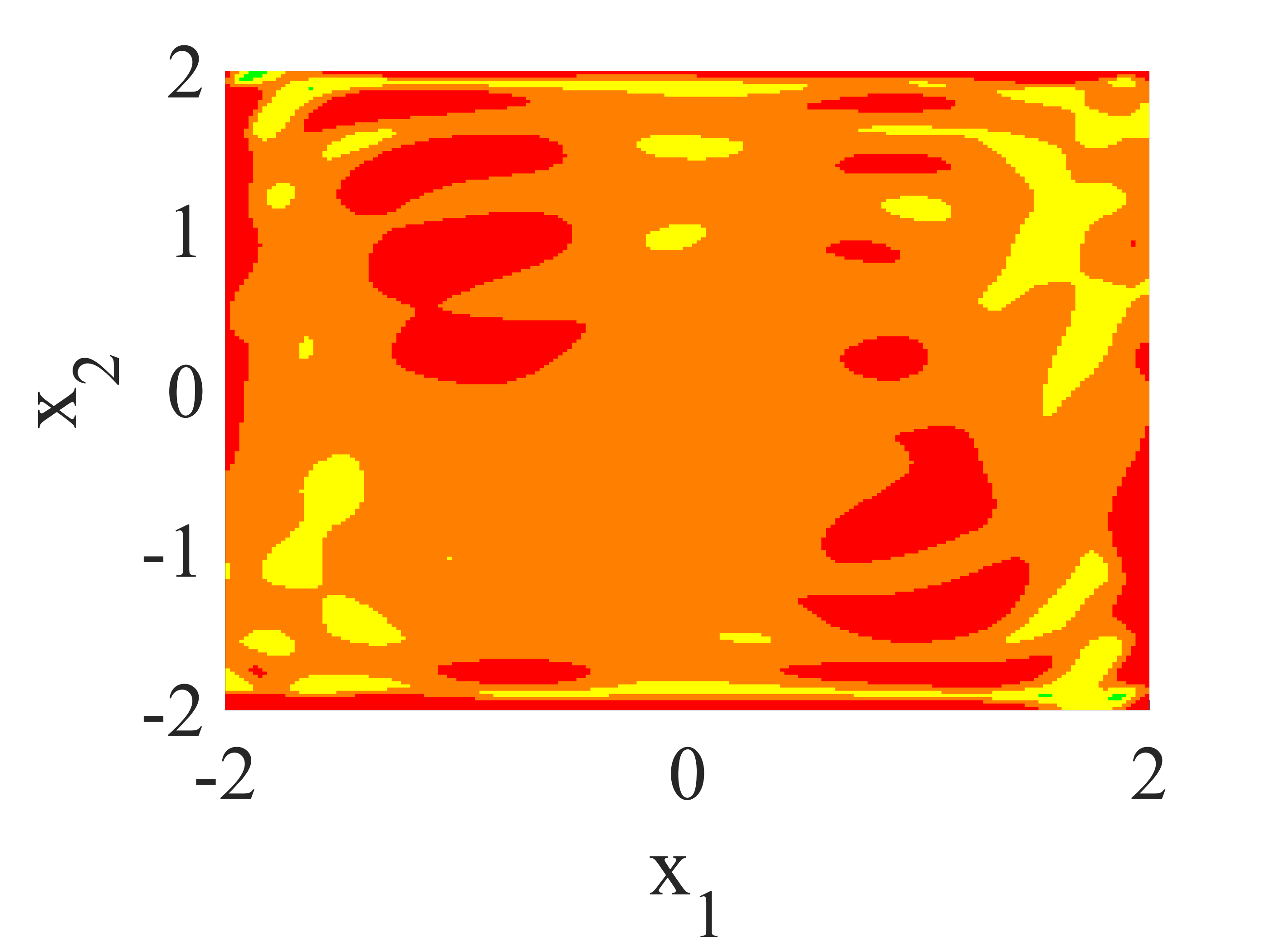}}
  \\
  {\includegraphics[width=.98\linewidth]{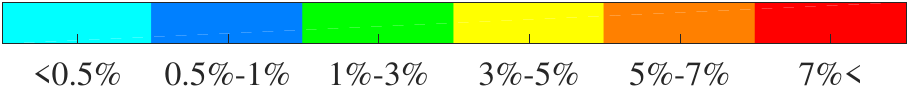}}
  \caption{Relative linear prediction error for dictionary identified by T-SSD ($\epsilon = 0.02$)
    (left) and the original dictionary (right) for~\eqref{eq:Duffing}.}\label{fig:relative_error_Duffing}
  \vspace*{-1ex}
\end{figure}

To illustrate the efficacy of our algorithm regarding the prediction
accuracy of the dictionary, Figure~\ref{fig:relative_error_Duffing}
compares the relative linear prediction
error~\eqref{eq:relative-linear-prediction-error} on the state
space~$\Mc$ for the dictionary $\dtssd$ identified by T-SSD with
$\epsilon =0.02$ and for the original dictionary~$D$ evaluated at
out-of-sample points other
than~$X$. Figure~\ref{fig:relative_error_Duffing} clearly shows
  the effectiveness of the T-SSD algorithm in improving the prediction
  accuracy.

To analyze the error for individual functions in the identified
  subspaces by T-SSD, we form a random test data set $\Xtest$,
  $\Ytest$ gathered with the same number of elements and sampling
  strategy used for~$X$
  and~$Y$. Table~\ref{table:RRMSEmax-vs-epsilon-Duffing} shows the
  maximum relative root mean square error defined
  in~\eqref{eq:rrmse_max} for the subspaces identified by T-SSD given
  different values of $\epsilon$. Despite the fact that we use
  different data for identification and evaluation,
  Table~\ref{table:RRMSEmax-vs-epsilon-Duffing} shows the
  effectiveness of the T-SSD algorithm for identifying subspaces on
  which all functions have prediction errors characterized by the
  accuracy parameter~$\epsilon$. 

{
	
	\renewcommand{\arraystretch}{1.5}
	\begin{table}[htb]
		\centering
		\caption{Maximum Relative Root Mean Square Error vs $\epsilon$ for~\eqref{eq:Duffing}.} \label{table:RRMSEmax-vs-epsilon-Duffing}
		
			\begin{tabular}[htb]{ | c || c | c | c | c | c | c| }
				\hline
				\textbf{$\boldsymbol{\epsilon}$}	               				    & 0.01  & 0.02  & 0.08  & 0.14  & 0.20 & 0.26  \\ \hline %
				\textbf{$\operatorname{RRMSE_{\max}}$}           & $\sim 0$        & 0.004       & 0.054  & 0.123    & 0.190 & 0.236    \\ \hline %
			\end{tabular}
		
	\end{table}
}

\subsection{Consensus on Harmonic Mean}
Given $N_a$ agents with state $x = [x_1, \ldots, x_{N_a}]^T $
communicating through a graph with adjacency matrix $A$, consider the
dynamics
\begin{align}\label{eq:Harmonic-mean-consensus}
  \dot{x_i} & = N_a \, x_i^2 \, \Upsilon(x)^{-2} \sum_{j=1}^{N_a}
  a_{ij} (x_j-x_i),\; i \in \until{N_a},
\end{align}
where $a_{ij}$ is the element of $A$ on row $i$ and column $j$ and
$\Upsilon(x)$ is the harmonic mean of the state elements defined as
\begin{align*}
  \Upsilon(x) = N_a \Big(\sum_{k=1}^{N_a} x_k^{-1} \Big)^{-1}.
\end{align*}
For any initial condition $x_0$, all the state elements converge to
the harmonic mean of the initial condition
$\Upsilon(x_0)$~\citep[Proposition~10]{JC:08-auto}, i.e., the agents
achieve consensus on $\Upsilon(x_0)$. For the purpose of this example,
we consider $N_a = 5$ agents communicating through an undirected ring
graph and states belonging to the state space $\Mc = [1,5]^5$. We
consider the discretized version of~\eqref{eq:Harmonic-mean-consensus}
with time step $\Delta t= 0.01s$ and gather $N =4 \times 10^4$ data
snapshots in matrices $X$ and $Y$ from $2 \times 10^4$ trajectories
with length equal to two time steps and initial conditions uniformly
selected from~$\Mc$. For our dictionary, we consider the space of all
polynomials up to degree $6$ and choose a dictionary $D$ with $N_d =
462$ functions spanning the space such that the columns of $\dx$ are
orthonormal.

We apply the Efficient T-SSD algorithm,
cf.~Section~\ref{sec:efficient_TSSD}, with $\epsilon \in \{0.05, 0.15,
0.3, 0.55,
0.8\}$. Table~\ref{table:dimension-vs-epsilon-Harmonic-mean-consensus}
shows the dimension of the identified dictionary, $\dtssd$, versus the
value of the design parameter~$\epsilon$. For $\epsilon = 0.8$, T-SSD
identifies the original subspace, certifying that the range spaces of
$\dx$ and $\dy$ are $0.8$-apart. On the other hand, the
one-dimensional subspace identified by $\epsilon = 0.05$ is in fact
the maximal Koopman-invariant subspace of $\Span(D)$, spanned by the
trivial eigenfunction $\phi(x) \equiv 1$ with eigenvalue $\lambda =1$.

{  \renewcommand{\arraystretch}{1.5}
  \begin{table}[htb]
    \centering
    \caption{Dimension of subspace identified by Efficient T-SSD vs
        $\epsilon$ for~\eqref{eq:Harmonic-mean-consensus}.} \label{table:dimension-vs-epsilon-Harmonic-mean-consensus} 
    
      \begin{tabular}[htb]{ | c || c | c | c | c | c |  }
        \hline
        \textbf{$\boldsymbol{\epsilon}$}	               				   & 0.05 & 0.15  & 0.30  & 0.55  & 0.80  \\ \hline %
        \textbf{$\mathbf{\boldsymbol{\dim \dtssd}}$}           & 1        & 14        &  64    & 272     & 462      \\ \hline %
      \end{tabular}
    
  \end{table}
}

\begin{figure}[htb]
  \centering 
  {\includegraphics[width=.90\linewidth]{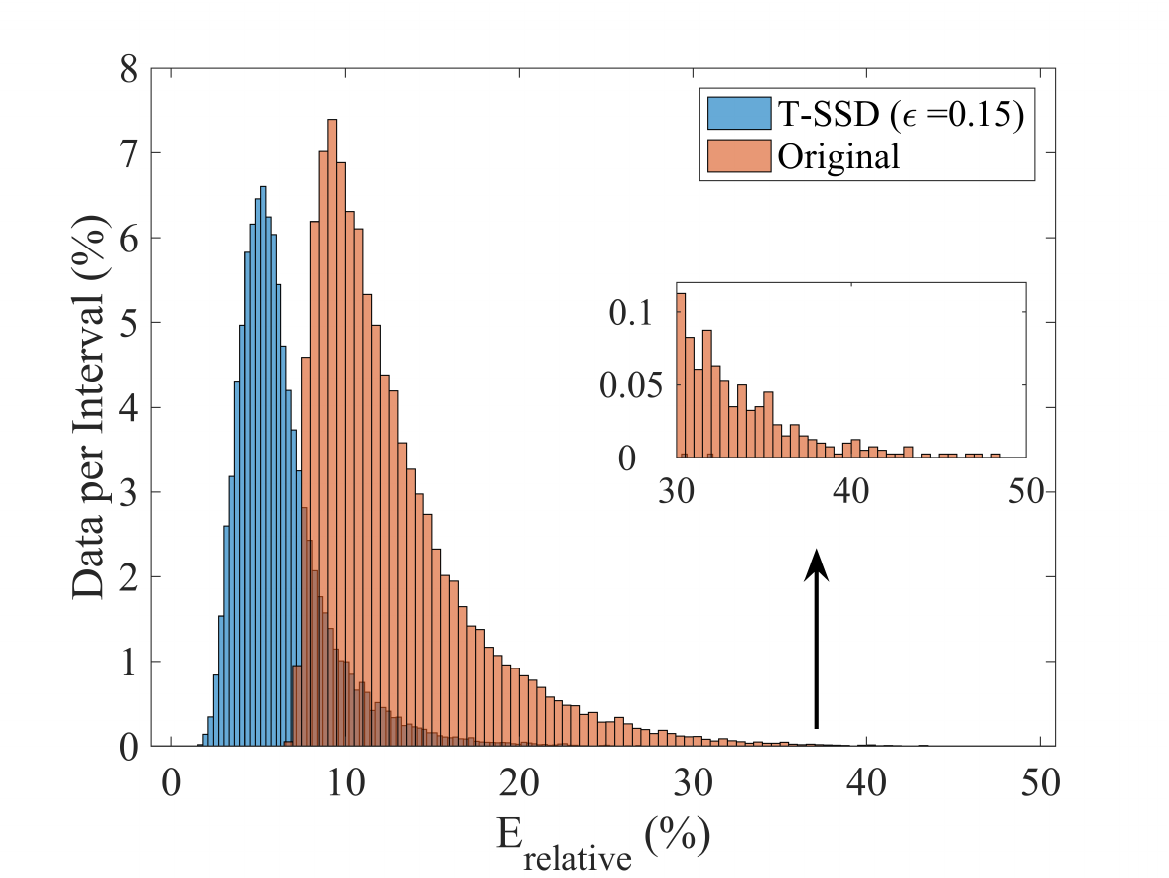}}
  \caption{Relative linear prediction error on test data for the
  		dictionary identified with T-SSD ($\epsilon = 0.15$) and
  		the original dictionary.}\label{fig:relative-linear-error-Harmonic-mean-consensus}
  \vspace*{-1ex}
\end{figure}

To illustrate the efficacy of T-SSD algorithm regarding prediction
accuracy, we first form a test data set comprised of snapshots
matrices $\Xtest$ and $\Ytest$ sampled with the same sampling strategy and
number of samples as $X$ and
$Y$. Figure~\ref{fig:relative-linear-error-Harmonic-mean-consensus}
provides histogram plots comparing the relative prediction error
defined in~\eqref{eq:relative-linear-prediction-error} of the
dictionary identified by T-SSD with $\epsilon = 0.15$ and the original
dictionary applied on the test data. In
Figure~\ref{fig:relative-linear-error-Harmonic-mean-consensus} the
horizontal axis denotes the prediction error while the vertical axis
shows the percentage of test data per
interval. Figure~\ref{fig:relative-linear-error-Harmonic-mean-consensus}
clearly shows the effectiveness of the T-SSD algorithm in improving
the prediction accuracy.

We also consider the prediction accuracy for the individual
functions. Table~\ref{table:RRMSEmax-vs-epsilon-Harmonic-mean-consensus}
shows the maximum relative root mean square error defined
in~\eqref{eq:rrmse_max} for the subspaces identified by T-SSD given
different values of $\epsilon$. According to
Table~\ref{table:RRMSEmax-vs-epsilon-Harmonic-mean-consensus}, despite
using different data for identification and evaluation, the error on
the test data satisfies the upper bound accuracy requirement enforced
by the accuracy parameter~$\epsilon$.

{  
  \renewcommand{\arraystretch}{1.5}
  \begin{table}[htb]
    \centering
    \caption{Maximum Relative Root Mean Square Error vs $\epsilon$ for~\eqref{eq:Harmonic-mean-consensus}.} \label{table:RRMSEmax-vs-epsilon-Harmonic-mean-consensus}
    
      \begin{tabular}[htb]{ | c || c | c | c | c | c | }
        \hline
        \textbf{$\boldsymbol{\epsilon}$}	               				   & 0.05  & 0.15  & 0.30  & 0.55  & 0.8  \\ \hline %
        \textbf{$\operatorname{RRMSE_{\max}}$}           & $\sim 0$        & 0.144       & 0.295 & 0.549    & 0.769     \\ \hline %
      \end{tabular}
    
  \end{table}
}

\section{Conclusions}\label{sec:conclusions}
We have presented the T-SSD algorithm, a data-driven strategy that
employs data snapshots from an unknown dynamical system to refine a
given dictionary of functions, yielding a subspace close to being
invariant under the Koopman operator. A design parameter allows to
balance the prediction accuracy and expressiveness of the algorithms'
output, which always contains the maximal Koopman-invariant subspace
and all Koopman eigenfunctions in the span of the original dictionary.
The proposed algorithm generalizes both Extended Dynamic Mode
Decomposition and Symmetric Subspace Decomposition.  Future work will
investigate noise-resilient strategies to approximate
Koopman-invariant subspaces and methods to construct expressive
dictionaries with high accuracy by alternating between growing the set
of functions (using specific basic functions or neural networks) and
pruning the dictionary to enhance accuracy while providing accuracy
bounds for all members of the identified vector space of functions.
Moreover, we aim to explore the application of the proposed method in
stability analysis, data-driven construction of Lyapunov
functions\footnote{The recent work~\citep[Section V.C]{GM-IA-TDM:22}
  provides an interesting example of using the T-SSD algorithm as a
  subroutine precisely for this purpose.}, and designing control
schemes with formal performance and stability guarantees by using the
Koopman operator to model control systems as bilinear or
switched-linear systems.

\appendix
\section{Basic Algebraic Results}\label{app1}

Here we collect two algebraic results from~\citep{MH-JC:22-tac} that
are used in our technical treatment.

\begin{lemma}\longthmtitle{\citep[Lemma
    A.1]{MH-JC:22-tac}}\label{l:subspace-intersection}
  Let $A, B \in \real^{m \times n}$ be matrices with full column rank.
  Suppose that the columns of $Z=[(Z^A)^T,(Z^B)^T]^T \in \real^{2n
    \times l}$ form a basis for the null space of $[A,B]$, where
  $Z^A,Z^B \in \real^{n \times l}$. Then,
  \begin{enumerate}
  \item $\range(AZ^A) = \range(A) \cap \range(B)$;
  \item $Z^A$ and $Z^B$ have full column rank.
  \end{enumerate}
\end{lemma}

\begin{lemma}\longthmtitle{\citep[Lemma A.2]{MH-JC:22-tac}}\label{l:product-subspace}
  Let $A,C,D$ be matrices of appropriate sizes, with $A$ having full
  column rank. Then $\range(AC) \subseteq \range(AD)$ if and only if
  $\range(C) \subseteq \range(D)$.
\end{lemma}

\end{document}